# Mobility Management in Emerging Ultra-Dense Cellular Networks: A Survey, Outlook, and Future Research Directions

**Syed Muhammad Asad Zaidi[1], Marvin Manalastas[1], Hasan Farooq[1] and Ali Imran[1], Senior Member, IEEE**

[1]University of Oklahoma, Tulsa, OK 74135 USA

Corresponding author: Syed Muhammad Asad Zaidi (e-mail: asad@ou.edu).

This work is supported by the National Science Foundation under Grant Numbers 1718956 and 1559483 and Qatar National Research Fund (QNRF) under Grant No. NPRP12-S 0311-190302. The statements made herein are solely the responsibility of the authors. For more details about these projects please visit: http://www.ai4networks.com.

**ABSTRACT** The exponential rise in mobile traffic originating from mobile devices highlights the need for making mobility management in future networks even more efficient and seamless than ever before. Ultra-Dense Cellular Network vision consisting of cells of varying sizes with conventional and mmWave bands is being perceived as the panacea for the eminent capacity crunch. However, mobility challenges in an ultra-dense heterogeneous network with motley of high frequency and mmWave band cells will be unprecedented due to plurality of handover instances, and the resulting signaling overhead and data interruptions for miscellany of devices. Similarly, issues like user tracking and cell discovery for mmWave with narrow beams need to be addressed before the ambitious gains of emerging mobile networks can be realized. Mobility challenges are further highlighted when considering the 5G deliverables of multi-Gbps wireless connectivity, <1ms latency and support for devices moving at maximum speed of 500km/h, to name a few. Despite its significance, few mobility surveys exist with the majority focused on adhoc networks. This paper is the first to provide a comprehensive survey on the panorama of mobility challenges in the emerging ultra-dense mobile networks. We not only present a detailed tutorial on 5G mobility approaches and highlight key mobility risks of legacy networks, but also review key findings from recent studies and highlight the technical challenges and potential opportunities related to mobility from the perspective of emerging ultra-dense cellular networks.

## I. INTRODUCTION

The unprecedented rise in the Internet traffic volume seen in recent years is attributed to high speed internet, and the advent of smart phone technology. It is anticipated that the number of 5G subscriptions will be 2.8 billion by the year 2025 [1]. Furthermore, the insatiable demand for new bandwidth-hungry applications will lead to an avalanche of traffic volume growth. Mobile data traffic will increase from 10.7 exabytes/month in 2016 to 83.6 exabytes/month by 2021 [2], and that number will further increase exponentially in the years to follow.

The emerging cellular networks including 5G mobile network standard as the next revolution of mobile cellular technology needs to support the ever-increasing mobile users, provide adequate data rate for the bandwidth hungry applications, address the QoS issues of delay tolerant applications and realize the concept of Internet-of-Things (IoT) [3] [4]. 5G promises to deliver “more” of everything [5]: a) top speeds of up to 1 Gbps, b) 100 Mbps data rate per end user even at the cell edge, c) RTT (Round-Trip-Time) latencies in the millisecond range, d) higher connection densities (1 million connections per $km^2$ [6]), and e) support for mobile devices at the speed of up to 500 km/h.

Currently, Signal to Interference and Noise Ratio (SINR) is considered as the primary metric for planning, dimensioning and optimization of the existing cellular networks [3]. However, for a few exceptions like fixed IoT services, an additional network planning/design criterion in the future may be the mobility related QoE. This is likely the outlook in the backdrop of the following observations:

1) Coverage and SINR provisioning will become a relatively easy challenge given the anticipated higher Base Station (BS) density in emerging cellular networks, along with the sophisticated interference management schemes and massive MIMO assisted beamforming.
2) However, the very same advances in the network design i.e. densification, beamforming, massive MIMO make the mobility management a more challenging problem. The challenges stem not only from the increased number of handovers (HOs) but also, beam management to maintain the expected QoE. Challenges related to beam management includes focusing narrow beams on the mobile users, cell discovery in narrow beam cells, and large signaling overheads when the user moves from one massive MIMO cell to another cell.
3) With the advent of mmWave, narrow beams of mmWave bands will have limited overlap with each other, making HO a challenging problem (see Fig.4 for observing the difference in HO scenarios in low frequencies and mmWave frequencies).

The growing demand for mobile services in public transport, highways, open-air gatherings etc. [7] will be critical to customer experience. Providing a satisfactory Quality of Experience (QoE) to a relatively large number of mobile users and a miscellany of the devices including phones, tablets, sensors etc. at the speed up to 500km/h imposes extreme challenges to the future mobile networks. Mobility requirements in emerging cellular networks require high efficiency of the HO mechanism, which makes the cell-change seamless for the users. Unlike the legacy technologies (i.e. 3G and 4G) that do not give primary importance to high mobility, future mobile networks will treat mobility as an integral part of the communication standard. Moreover, the mobility management schemes in Long Term Evolution (LTE) systems (also known as 4G system) and to a certain extent, even in the latest 5G New Radio (NR) standard are not well adapted to the typical deployment of the futuristic mobile networks due to multiple factors, few of which are highlighted below:

- The legacy LTE architecture makes use of a centralized network control entity called MME (Mobility Management Entity) located in the core network. The emerging cellular networks are expected to have 10-folds higher density [8], with a larger fraction of mobile users. Thus, without a mobility centric redesign of the architecture, future networks should have 10 times more MME's just to achieve a similar QoS as in LTE.
- To achieve the logistic feasibility for high density deployment, BS placement in future mobile networks are likely to be impromptu or much less planned [8]. This will increase mobility related signaling load that is bound to complicate the core network management and planning.
- HO decision in existing networks is made by participating BSs without considering the deployment of the BSs and backhaul limitations. In futuristic mobile networks with flexible BS deployment, the chances of User Equipment (UE) in selecting the optimal target BS may become smaller.
- While the capacity crunch will be addressed by small-cells (SC), a large number of inter-SC HOs will take place leading to frequent session interruptions during HO.
- With smaller inter-site-distance as in SCs, the performance of the existing mobile network reduces sharply owing to the risk of HO failures due to high radio link variability as shown in [9].
- In existing mobile networks, UE context has to move from one BS to another for every HO. This will impose unprecedented signaling overhead in the future ultra-dense network architecture. While signaling is already growing 50% faster than data traffic [10], network efficiency will drop by many folds using the current HO approaches.
- HOs in 4G networks are based on the broadcast signal called Reference Signal (RS). The mmWaves with narrow beams cannot have RS broadcast to the whole coverage area within the cell range. Hence, cell discovery, especially for mobile UEs is another key mobility challenge in emerging cellular networks not faced by the traditional mobile networks.
- With SON stepping up the automatization of network configuration and optimization in LTE, myriad of mobility management parameters associated with the large number of closely deployed 5G BSs need to be well managed. For that, the existing SON solutions will not be sufficient.
- 5G applications with Ultra Reliable Low Latency Communications (URLLC) e.g. self-driven cars demand very low latency requirements as shown in Table I [11].
- When UE perform HO to a better cell, it experiences a latency and data interruption period. HO management in the future mobile networks should ensure a seamless and latency-free transition from the source to the target cell.
- With mobile phone traffic on the rise, and with the advent of self-driven cars and drones needing robust connectivity, seamless and reliable mobility management has become more significant than ever. The adaptation of ultra-dense cellular networks and mmWave BSs makes the mobility management even more complex challenge requiring significant research effort.

TABLE I
COMPARISON OF LTE LATENCY WITH 5G EXPECTED GOALS

| Parameter | LTE Requirement | 5G Target |
|---|---|---|
| Control Plane Latency (Accessibility) | 100ms | 10ms |
| User Plane Latency | 20ms | 1ms |
| HO Execution | 49.5ms | 0ms |

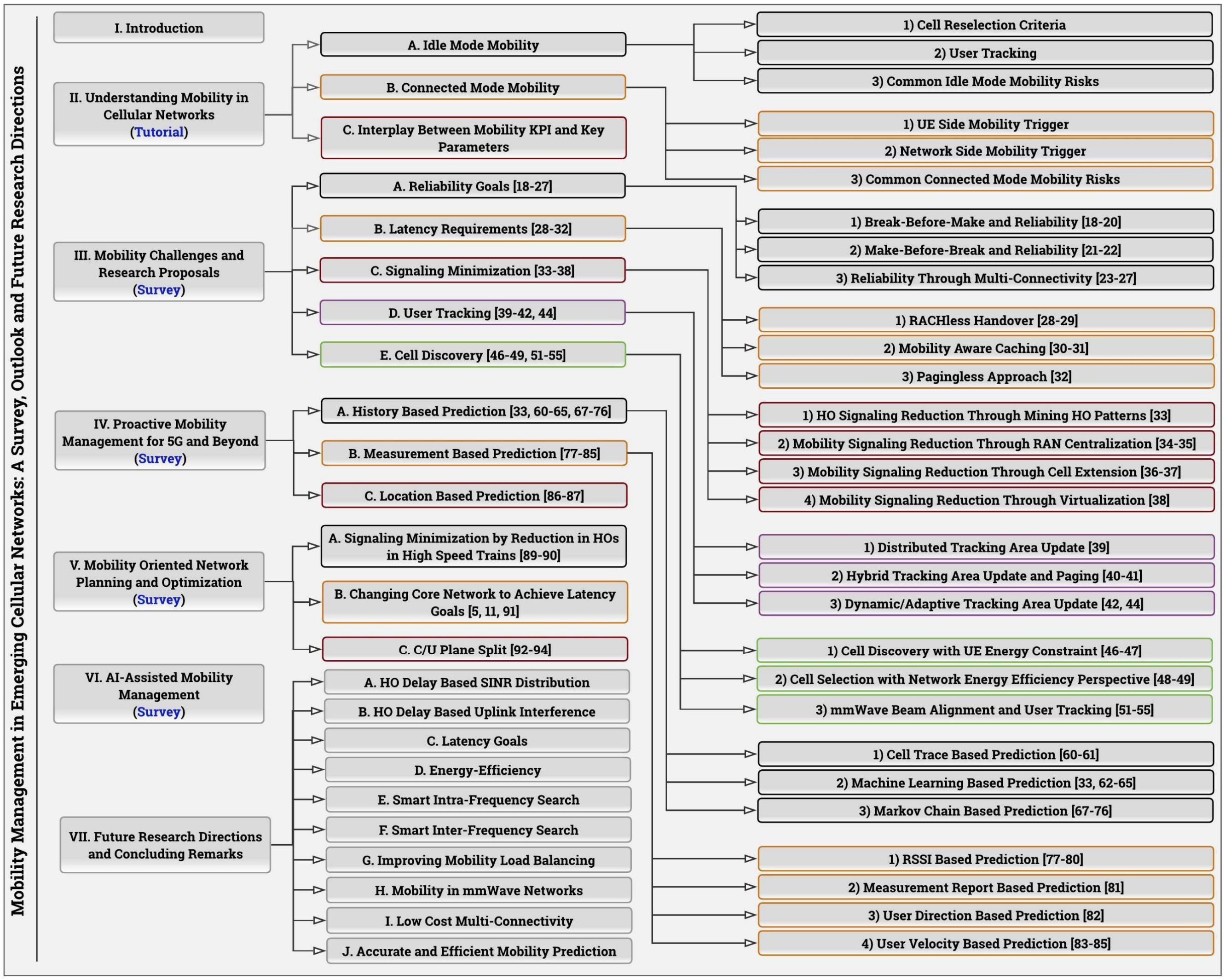

FIGURE 1. **Layout of the contents and paper contributions.**

In light of the above discussion, we can conclude that mobility management will have much stronger impact on the design and architecture of upcoming cellular networks, than it had on the legacy networks. The futuristic networks will incorporate high mobility requirements as an integral part, and appreciable efforts are required to attain ubiquitous top-notch QoE.

Majority of mobility oriented surveys in the literature target adhoc networks [12] [13] [14]. Mobility surveys on cellular networks do exists e.g. Xenakis et al. [15] presented survey on HO decision algorithms for the femtocells in LTE-Advance. Another survey on high mobility wireless communication has recently been presented in [16], however, the attributes and intricacies of the 5G architecture have not been addressed. To the best of the authors' knowledge, this survey is the first to address the novel contributions by research community targeting mobility in emerging ultra-dense mobile networks. The contributions in this paper and its organization are as follows:

- To the best of the authors' knowledge, this paper gives the first comprehensive tutorial on 3GPP based 5G mobility management procedures for both a) idle/inactive mode, and b) connected mode mobile users.
- Mobility related surveys do exist in the literature (e.g. [12] [13] [14] on adhoc networks), but none of the aforementioned surveys addresses the futuristic mobile networks. This paper presents a single go-to manuscript where future researchers not only understand the 3GPP mobility procedure and the existing mobility related literature but also assist them in finding the research directions they might undertake.
- It presents a first of its kind framework to correlate all mobility management related parameters with all mobility management related KPIs. To facilitate easy understanding, this framework is presented in the form of a flow chart shown in Fig. 8.
- It presents a comprehensive and taxonomized review of the literature on mobility management.

- It identifies the need for a new paradigm for mobility management deemed essential to meet the quality of experience (QoE) requirements of the emerging applications and use-cases.
- It proposes a novel proactive mobility management framework to meet the requirements of the emerging mobile networks. Since the challenges of 5G networks (e.g. low latency, less overhead and high quality of experience) cannot be addressed by the current reactive mobility management techniques, we discussed the proactive mobility management in section IV.
- It highlights the need to come up with Mobility oriented Network planning and dimensioning
- It provides a collection of the latest AI-based techniques to smartly address mobility related challenges.
- It identifies the future research direction and few open research problems to achieve this paradigm shift.

Fig. 1 outlines the structure of the paper. It also provides a taxonomy of the literature on mobility.

## II. UNDERSTANDING MOBILITY IN CELLULAR NETWORKS

Mobility in cellular networks plays a pivotal role ensuring an optimal experience to the subscribers. It guarantees that mobile users won't just be able to maintain connectivity but attain the best available connection to the network as they move towards the destination. Seamless and timely HO and cell reselection has always been a major challenge in any wireless communication systems including 5G. Mobility has been categorized as Idle and Connected Mode Mobility in 5G. Note that the mobility procedure in LTE (4G) is very similar in 5G New Radio (NR) using events A1, A2, A3, A4, A5 and A6 to trigger HOs. Event A2 and A1 are triggered when RF condition of the UE falls below and exceeds the configured threshold respectively and are used to start and stop inter-frequency neighbor search. Intra-frequency HO is initiated by event A3 where the neighbor RF condition becomes higher than serving RF condition by a configured threshold. Event A4 and A5 are typically used for inter-frequency HO where target inter-frequency cell has to be higher than an absolute threshold for the event A4 to be triggered. On the contrary, event A5 in addition to event A4 condition, requires serving cell RF condition to be below a certain threshold. Finally, event A6 is similar to event A3 but is used for intra-frequency HO of the secondary frequency the UE is camped onto. Event A4 and A5 can also be used for conditional HO management for e.g. for load balancing. In addition to the events described above, event B1 and B2 (A4 and A5 alike) are also used for inter-technology HO, and for dual-connectivity, but they are not discussed here to keep the focus of this paper confined to basic mobility procedures and the associated challenges.

The only difference between 5G and 4G mobility criteria is in the idle mode where respective idle mode reselection parameters in 5G NR are present in different SIB# than in LTE. Moreover, the idle mode parameter names and functionalities in 5G are similar as in 4G. Comprehensive explanation of 5G mobility procedure while keeping in view the 5G network architecture and interfaces is presented in the following subsections.

TABLE II
LIST OF ACRONYMS

| Acronym | Description |
|---|---|
| 3GPP | Third Generation Partnership Project |
| 4G | Fourth Generation |
| 5G NR | Fifth Generation New Radio |
| AMF | Access & Mobility Function |
| BS | Base Station |
| CDR | Call Detail Record |
| CIO | Cell Individual Offset |
| CoMP | Co-Ordinated Multi Point |
| CQI | Channel Quality Indicator |
| CSI | Channel State Identifier |
| gNB | 5G Base Station (Next Generation NodeB) |
| HF | High Frequency |
| HO | Hand Over |
| HOM | Hand Over Margin |
| IMMCI | Idle Mode Mobility Control Information |
| ICIC | Inter Cell Interference Coordination |
| IoT | Internet of Things |
| KPI | Key Performance Indicator |
| LB | Load Balancing |
| LoS | Line of Sight |
| LTE | Long Term Evolution (4G) |
| MLB | Mobility Load Balancing |
| MME | Mobility Management Entity |
| MR | Measurement Report |
| MRO | Mobility Robustness Optimization |
| MIMO | Multiple Input Multiple Output |
| MDT | Minimization of Drive Test |
| NLoS | Non-Line of Sight |
| PCI | Physical Cell Identifier |
| P-GW | PDN Gateway |
| QoE | Quality of Experience |
| RAT | Random Access Technology |
| RRC | Radio Resource Control |
| RTT | Round Trip Time |
| RS | Reference Signal |
| RSRP | Reference Signal Receive Power |
| RSRQ | Reference Signal Receive Quality |
| RSSI | Receive Signal Strength Indicator |
| RwR | Release with Redirect |
| RLF | Radio Link Failure |
| SC | Small Cell |
| SINR | Signal to Interference plus Noise Ratio |
| S-GW | Serving Gateway |
| SON | Self-Organizing Networks |
| SDN | Software Defined networking |
| SIB | System Information Base |
| TA | Tracking Area |
| TAL | Tracking Area List |
| TAU | Tracking Area Update |
| UPF | User Plane Function |
| UE | User Equipment |
| URLLC | Ultra-Reliable Low Latency Communication |
| UDN | Ultra-Dense Cellular Networks |

### A. IDLE MODE MOBILITY

UE is in idle mode when it is neither running any active communication service nor is connected to any particular cell. UE in idle mode is constantly trying to search and maintain services such as Public Land Mobile Network selection, cell selection and reselection, location registration, and reception of system information. By maintaining an idle mode connection, UE can readily establish a Radio Resource Connection (RRC) for signaling or data transfer as well as be able to receive any possible incoming connections.

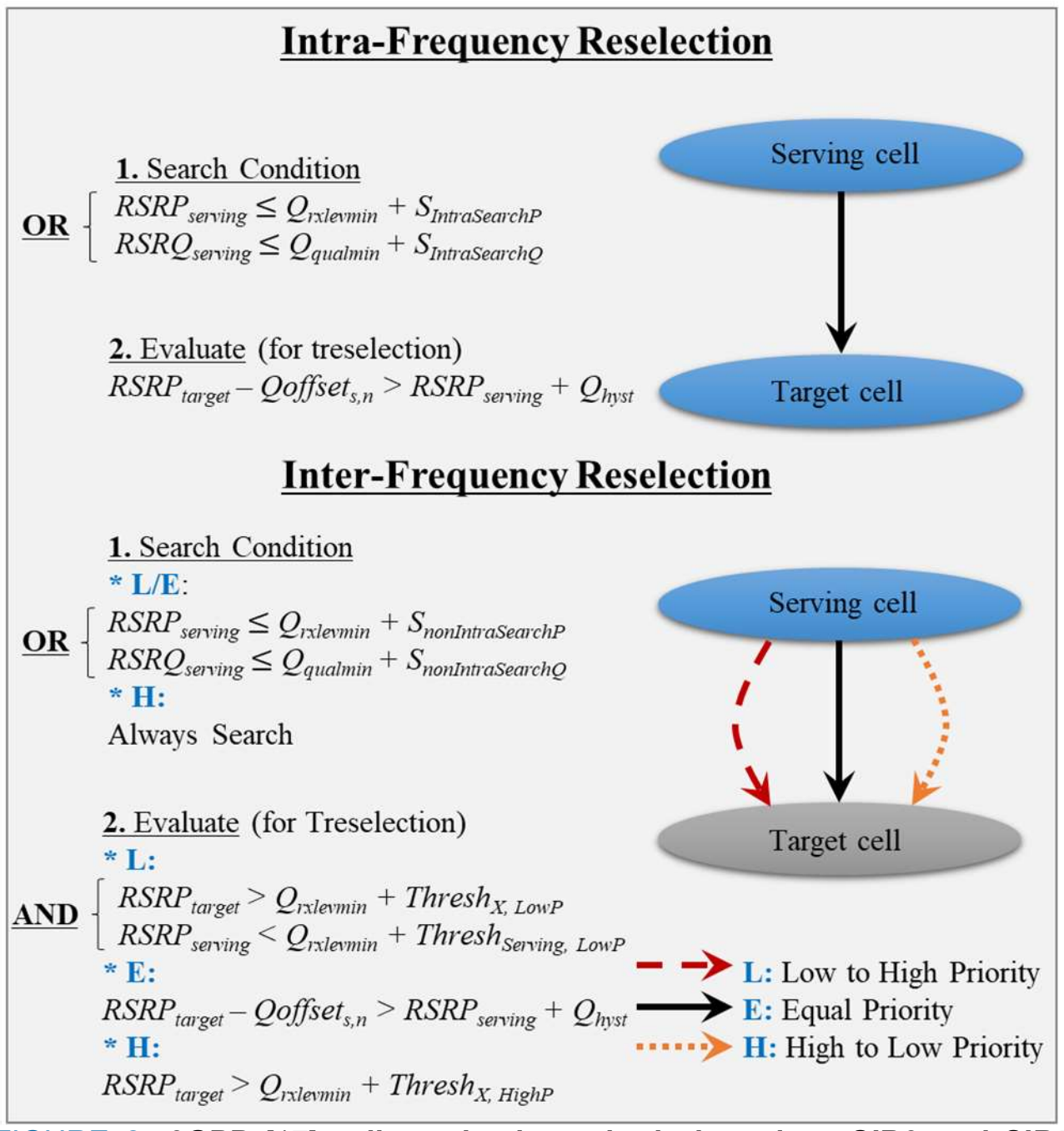

FIGURE 2. **3GPP [17] cell reselection criteria based on SIB3 and SIB5 parameter for intra-frequency and inter-frequency reselection respectively.**

UE always powers ON in idle mode and selects the cell with the maximum signal strength through a process known as cell selection. However, this initially selected cell will not always be the best to serve especially when UE moves from one place to another. Therefore, to maintain the quality of signal, UE has to camp on another optimal cell, a process known as cell reselection.

#### 1) CELL RESELECTION CRITERIA

In 5G, BS broadcasts nine System Information Block (SIB) messages for the UE as defined in 3GPP [17]. Out of those messages, SIB 1, 2, 3 and 4 contain critical parameters to execute idle mode cell reselection to the optimal 5G cell. SIB1 has the serving cell parameters as well as the cell selection parameters, while SIB2 has the common parameters used for intra-frequency and inter-frequency reselection. SIB3 is dedicated to intra-frequency reselection parameters, however, operators can broadcast the related parameters in SIB2 instead, and thus SIB3 is not broadcasted. SIB4 contains inter-frequency reselection through target frequency priority and the associated parameters. Fig. 3 illustrates a pictorial demonstration of the reselection conditions and evaluation in 5G as described by 3GPP. Description of the related reselection parameter, and the respective location (SIB#) can be found in Table III. LTE uses the same reselection procedure with the only difference that the contents of SIB2, SIB3 and SIB4 in 5G are found in SIB3, SIB4 and SIB5 of LTE instead.

#### 2) USER TRACKING

The idle mode mobility of the UE is the responsibility of Access and Mobility Function (AMF) at the Tracking Area (TA) level for RRC idle mode users and at the RAN Notification Area (RNA) for RRC inactive mode users. Here we only talk about the idle mode users as the mobility procedure in 5G is similar for RRC idle mode and RRC inactive mode users. Note that unlike the connected mode, network is unaware of cell-level UE location in idle mode. After powering ON, UE acquires the Tracking Area List (TAL) composed of a list of TA codes through the periodic SIB1 broadcast from the cell. As UE traverses through the network while performing cell reselection procedure, it compares the TA code of the new cell with its own TAL. If the TA code of a newly visited cell does not match with its own TAL, it initiates TA Update (TAU) process to request AMF for location update as seen in the Fig. 3(a). TAU helps to track the UE in case of any incoming call. Bigger TA size reduces signaling overhead of TAU at the expense of larger paging domain, ultimately resulting in higher paging-based downlink signaling load at network level.

TABLE III
3GPP [17] INTRA/INTER-FREQUENCY RESELECTION PARAMETERS

| Parameter | SIB# | Description |
|---|---|---|
| SIB | - | System Information Broadcast |
| RSRP | - | Reference Signal Received Power |
| RSRQ | - | Reference Signal Received Quality |
| $Q_{rxlevmin}$ | SIB1 | Minimum RSRP threshold required to camp in idle mode |
| $Q_{rxlevmin}$ | SIB2 | $RSRP_{serving}$ threshold required to compute intra-frequency reselection conditions |
| $Qoffset_{s,n}$ | SIB2 | Positive or negative bias required to promote or avoid intra-frequency cell reselection to target cell * *Idle Mode Cell Individual Offset* |
| $Q_{hyst}$ | SIB2 | $RSRP_{target} - RSRP_{serving}$ required to satisfy intra-frequency reselection condition. |
| *Treselection* | SIB2 | Time needed to satisfy intra-frequency reselection condition before actual reselection to the optimal cell |
| $S_{IntraSearchP/Q}$ | SIB2 | RSRP/RSRQ threshold below which user searches for intra-frequency target cell |
| $Q_{rxlevmin}$ | SIB4 | $RSRP_{serving}$ threshold required to compute inter-frequency reselection conditions |
| $Q_{qualmin}$ | SIB4 | $RSRQ_{serving}$ threshold required to compute inter-frequency reselection condition |
| $Qoffset_{s,n}$ | SIB4 | Positive or negative bias required to promote or avoid inter-frequency cell reselection to equal priority target cell * *Idle Mode Cell Individual Offset* |
| $Q_{hyst}$ | SIB4 | $RSRP_{target} - RSRP_{serving}$ required to satisfy reselection condition to equal priority cell |
| *Treselection* | SIB4 | Time needed to satisfy inter-frequency reselection condition before actual reselection to the optimal cell |
| $S_{nonIntraSearchP/Q}$ | SIB4 | $RSRP_{serving}$ / $RSRQ_{serving}$ threshold below which user searches for inter-frequency target cell |
| $Thresh_{X, LowP}$ | SIB4 | $RSRP_{target}$ threshold required to trigger inter-frequency reselection to lower priority target cell |
| $Thresh_{Serving, LowP}$ | SIB4 | $RSRP_{serving}$ threshold required to trigger inter-frequency reselection to lower priority target cell |
| $Thresh_{X, HighP}$ | SIB4 | $RSRP_{target}$ threshold required to trigger inter-frequency reselection to higher priority target cell |

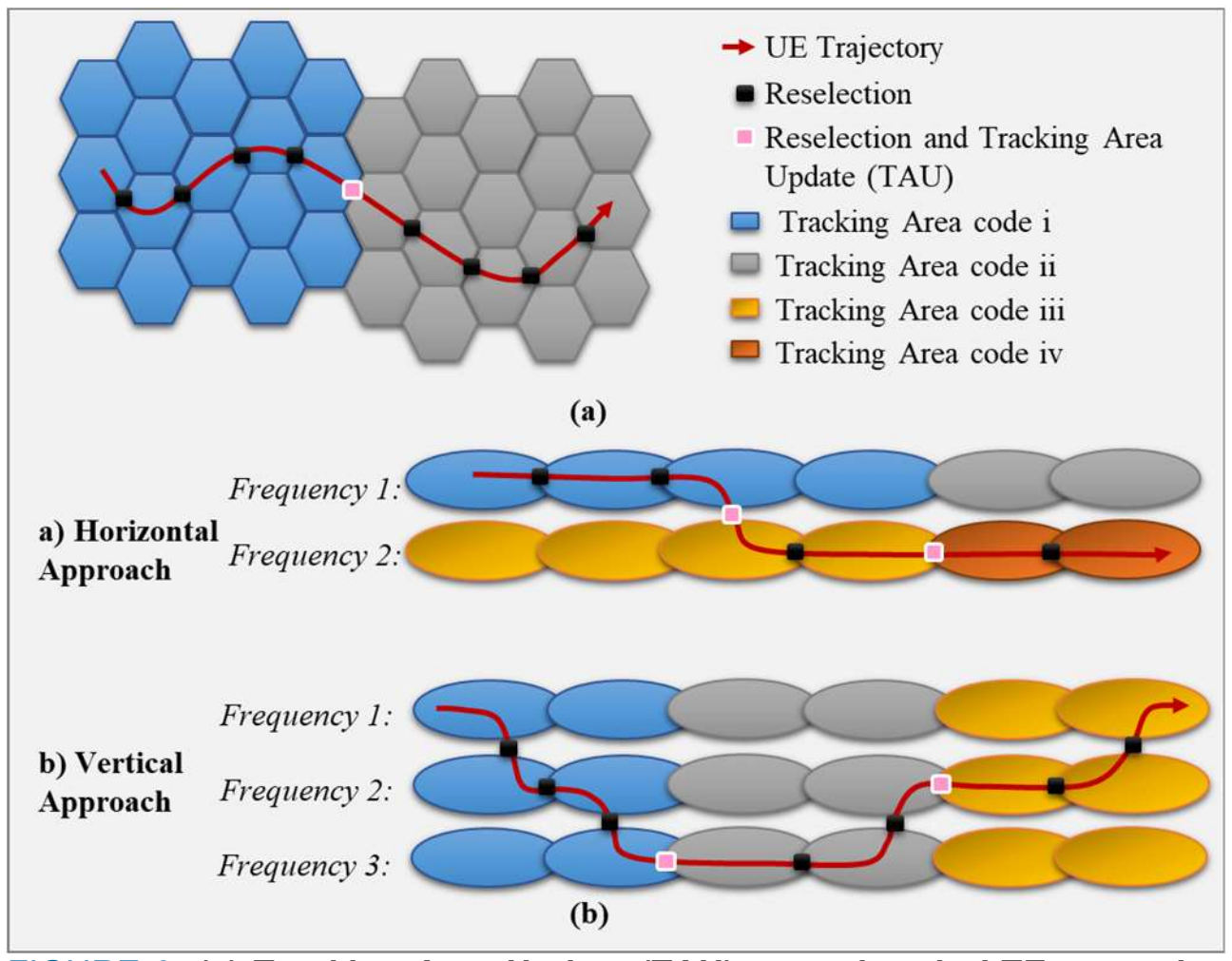

FIGURE 3. **(a) Tracking Area Update (TAU) procedure in LTE networks, (b) Common Tracking Area (TA) planning approaches.**

### 3) COMMON IDLE MODE MOBILITY RISKS

In this subsection, we discuss about the common idle mode mobility risks in the existing LTE network. But since the mobility process is similar in 5G networks, 5G capable UEs are expected to face similar challenges.

In idle mode, data transmission does not take place, therefore reliability and QoS are not the issues of concern. However, reselection procedure can incur accessibility and user tracking issues in rare occasions.

During the network attach procedure, idle mode UE first sends connection request and awaits connection setup message from the BS. If UE does not receive any message from the BS within a predefined time (t300 timer known to UE via SIB2 'SIB1 in 5G [18]'), it restarts the accessibility procedure. Under special circumstances, if UE sends a connection request to the serving cell followed by reselection to a neighboring cell, it cannot receive the connection grant simultaneously. The new serving cell in this case does not become aware that the UE which just moved under its coverage needs to access the network. Thus, UE has to wait for a time defined in t300 before re-initiating the access procedure in the new serving cell. During this time, UE experiences latency and can have serious impact on the applications requiring ultra-low latency. The delay can be suppressed by having smaller t300 timer, but at the cost of increased signaling load due to the increase in redundant connection requests and replies. Moreover, smaller t300 also negatively impact UE energy consumption (due to recurrent Random-Access Channel 'RACH' attempts). Repeated RACH attempts might result in higher Central Processing Unit (CPU) load of serving cell, especially at busy hour.

Similar accessibility delay at TA border can result in paging failure, since the network can be unaware of the accurate UE location unless TAU followed by a successful accessibility is performed.

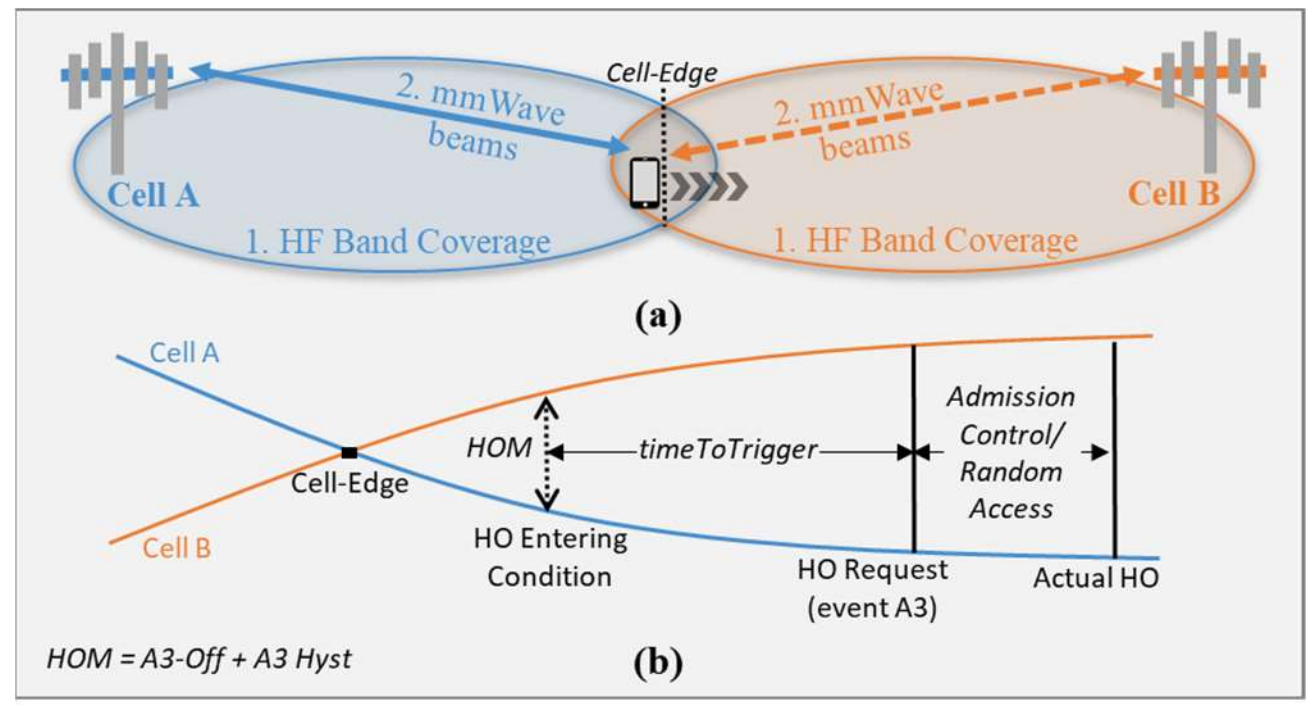

FIGURE 4. **General HO procedure. (a) UE performs HO from cell A to cell B at cell-edge as it moves closer to the cell B. Scenario 1 and 2 represents HF coverage and mmWave narrow beams, (b) 3GPP** [18] **based intra-frequency HO process.**

TA planning is a crucial task and two approaches are used in existing networks: a) horizontal approach, b) vertical approach, as shown in Fig. 3(b). TAU procedure initiates for every inter-frequency reselection in horizontal approach, thus it is deployed where radio condition is good, and user is least expected to make recurrent inter-frequency reselection. On the contrary, poor radio condition area should have vertical approach to minimize TAU for inter-frequency reselection instances. Horizontal approach is favorable for high speed traffic like train lines or highways. One approach to address this issue in the existing cellular network is the use of adaptive TA codes, where users are configured with a list of TA codes to prevent ping-pong TAUs. However, determining the optimal number of TA codes in a list and the cumulative TA size still remain an open research problem.

## B. CONNECTED MODE MOBILITY

UE is said to be in connected mode when it has established a connection with its peer Radio Resource Control (RRC) layer at the serving BS and the network can transmit and/or receive data to/from the UE. As there is an exchange of data between

TABLE IV
3GPP [18] HANDOVER PARAMETERS CONVEYED TO UE IN RRC RECONFIGURATION LAYER 3 MESSAGE

| Parameter | Description |
|---|---|
| *s-Measure* | RSRP threshold below which user searches for optimal intra-frequency target cell |
| *Ofn* | Frequency offset for target cell |
| *Ofp* | Frequency offset for serving cell |
| *Ocn* | Target cell offset<br>* *Commonly known as Cell Individual Offset 'CIO'* |
| *Ocp* | Serving cell offset |
| *Hys** | Hysteresis to prevent ping-pong HOs |
| *A3-Off** | $RSRP_{target} – RSRP_{serving}$ offset required to satisfy A3 condition |
| *A2-Thr** | Event A2 $RSRP_{serving}$ threshold |
| *A1-Thr** | Event A1 $RSRP_{serving}$ threshold |
| *A4-Thr** | Event A4 $RSRP_{serving}$ threshold |
| *A5-Thr1** | Event A5 $RSRP_{serving}$ threshold |
| *A5-Thr2** | Event A5 $RSRP_{target}$ threshold |
| *timeToTrigger (TTT)* | Time for which Event (A1-A5) condition need to be satisfied before sending measurement report to the Base Station |

* Combination of (A1/A2/A3/A4/A5)-Thr(s) and respective Hys parameter are used to define each event.

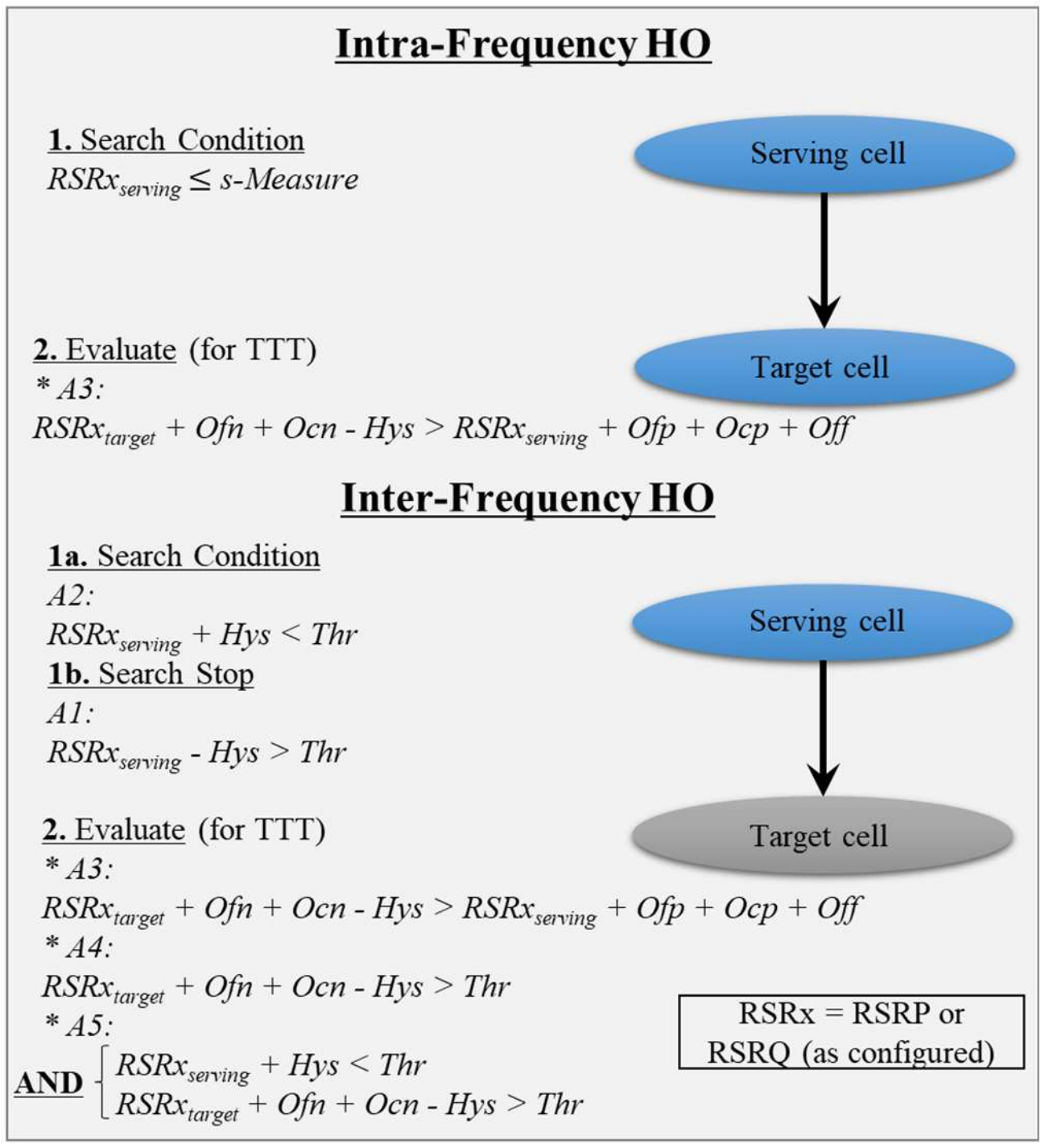

FIGURE 5. **3GPP [18] intra-frequency and inter-frequency handover criteria in LTE networks.**

the UE and the BS, uninterrupted data transfer needs to take place for a seamless continuity of service when a UE moves from one BS to another BS. This ideally seamless mobility in connected mode is termed as handover (HO).

### 1) UE SIDE MOBILITY TRIGGER

UE triggers an intra-frequency HO request to the next optimal cell by sending A3-Measurement Report (MR) to its serving cell as shown in Fig. 4. The serving cell then decides whether to entertain the request and perform the HO, by communicating with the target cell and serving AMF. An intra-frequency HO is the first preference in cellular networks; however, there are instances in which an inter-frequency HO is the preferred choice. For example: a) when there is a coverage hole in the serving frequency, b) when the current serving cell does not support the requested service e.g. Voice over NR, and c) when load balancing is needed to avoid congestion in the serving frequency. In Fig. 5 we illustrate the 3GPP [18] defined inter-frequency HO criteria. For a description of each HO parameter, refer to Table IV.

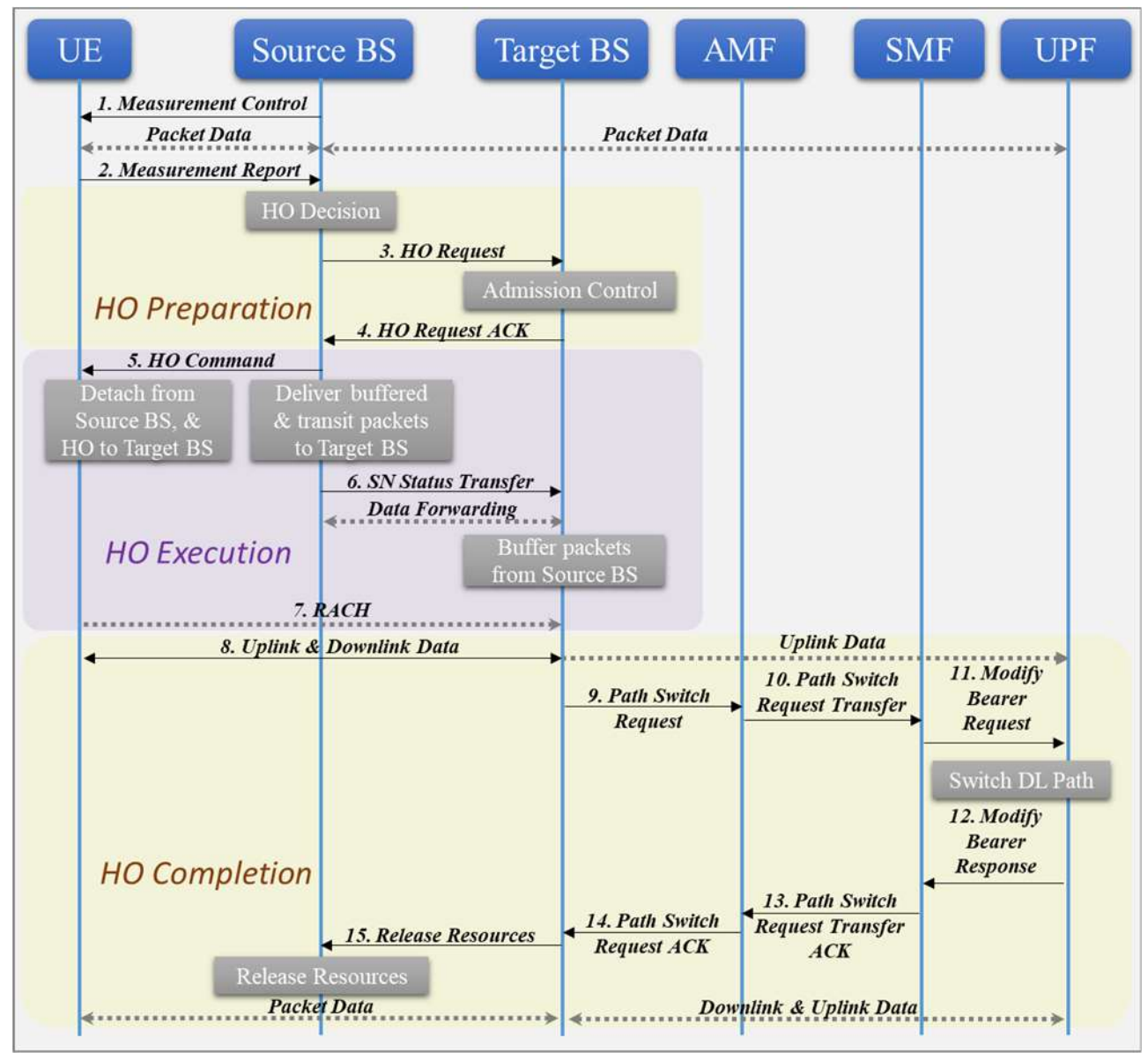

FIGURE 6. **Xn based handover without UPF re-allocation in 5G networks.**

### 2) NETWORK SIDE MOBILITY TRIGGER

HOs are undoubtedly more complicated than cell reselection. Aside from the source and target cell, core entities which include Access and Mobility Function (AMF), Session Management Function (SMF) and User Plane Function (UPF) need to be updated as well. Depending on the scenario, data transfer and handling could pose several challenges. In normal cases, when AMF, SMF and UPF do not change during the HO, signaling is reasonable and it is termed Xn based HO.

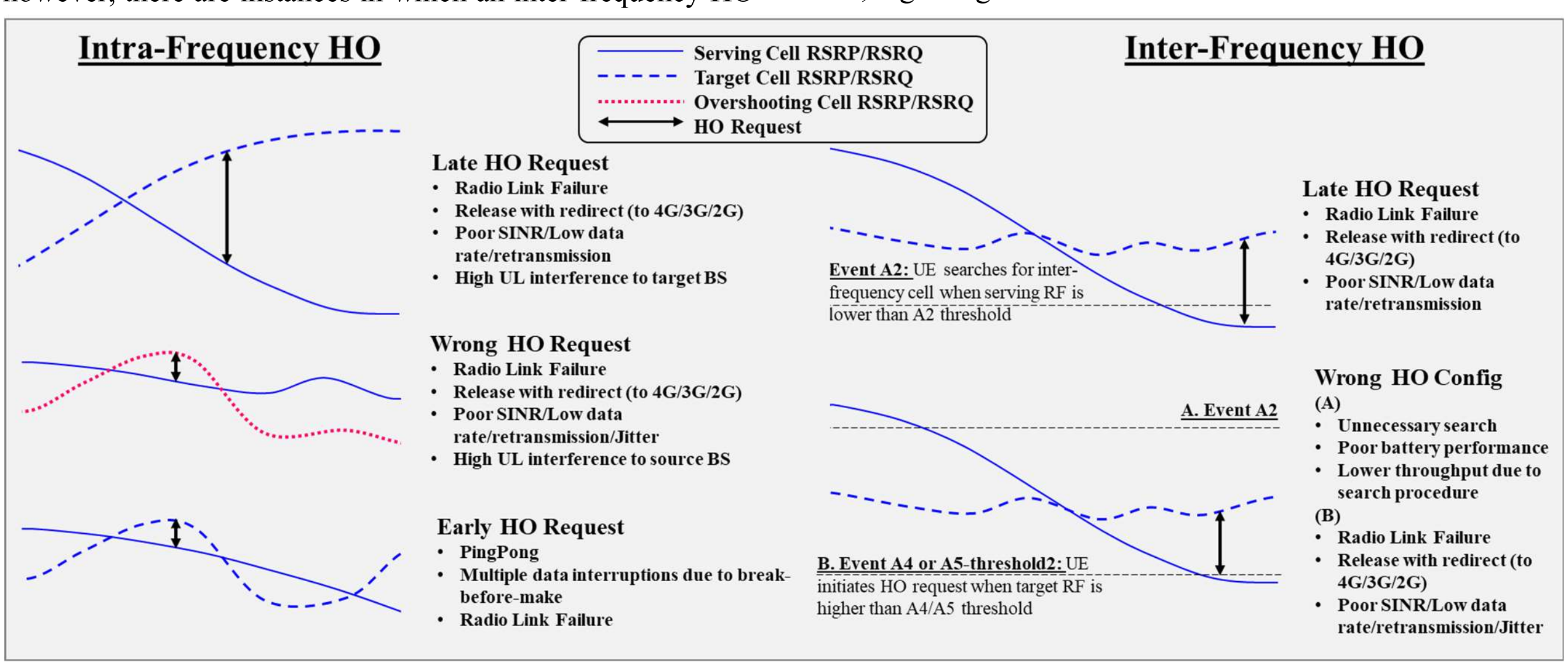

FIGURE 7. **Common Mobility Related Risks in 4G/5G networks.**

TABLE V
COMMON HO ISSUES AND THEIR SOLUTIONS

| HO Issue | Parameter Optimization Solution | Possible Cons |
|---|---|---|
| Late Intra HO | i. Lower A3 *offset*, shorter *TTT* | Prone to unwanted HO's to non-target cells. |
| | ii. Positive *CIO* towards target cell | Potential Ping-Pong between source and target especially for static users. |
| Late Inter HO | Higher A2, Accelerate A3/A4/A5, shorter *TTT* | Prone to unwanted HO's to non-target cells/layers. |
| Wrong Intra HO | i. Higher A3 *offset*, longer *TTT* | May cause delayed HO to target cell. |
| | ii. Negative *CIO* towards wrong-target cell | Stationary users might experience poor signal quality. |
| Wrong Inter HO | Lower A2, Delay A3/A4/A5, shorter *TTT* | May cause delayed HO to target cell. |
| Early Intra HO | i. Higher A3, longer *TTT* ii. Negative *CIO* towards target cell | May cause HO delay to target cell |
| Early Inter HO | Lower A2, Delay A3/A4/A5, shorter *TTT* | |

Abbreviations: *CIO* = Cell Individual Offset, *TTT* = timeToTrigger, Intra HO = Intra-frequency hand over, Inter HO: Inter-frequency hand over.

Here, the Xn interface is used for the preparation phase of the HO. However, when the Xn interface does not exist between the participating cells, an N2 based HO is performed where cells use a longer path for communication. Signaling flow for the Xn based HO is illustrated in Fig. 6. 3GPP [18] named Xn as the interface used to connect 5G BSs directly, and N2 interface is the logical interface between two 5G BSs connected through the core network (AMF). N2 interface is used if the direct Xn interface between the neighboring BSs do not exists.

#### 3) COMMON CONNECTED MODE MOBILITY RISKS

Apart from the fast fading effect due to Doppler shift in physical layer, the mobile UE has to cope with several Layer 3 issues as well, which can be eluded primarily by a timely HO and an optimal selection of the target BS. Some of the issues mobile UE experiences during inter-site mobility are presented in Fig. 7, with possible solution(s) in Table V.

### C. INTERPLAY BETWEEN MOBILITY KPI AND KEY PARAMETERS

Network operators optimize their network by tuning a set of mobility related parameters, and then by observing the HO attempt, HO success and few other QoE KPIs affected by those modified network parameters. Few of the vital mobility related KPIs are outlined below:

- User tracking KPI indicates the paging hit rate when users served under the TA are notified by an incoming call. The idle mode mobile user must update its location (via TAU) to the core network when it moves into the neighboring TA. By doing so, the respective TA is broadcasted with paging attempt messages in case of any incoming call. A delay in TAU can result in paging failure and reattempts.
- Mobility oriented HO process or TAU trigger results in the control plane messages being sent in the air interface and in the core network. The percentage of network resources used by control plane are measured by signaling data KPI.
- User terminal energy consumption e.g. during data delivery and location update, can be measured by the UE battery KPI.
- Reliability (or retainability) KPI indicates the percentage of users that dropped the connection with their participating cells during the HO procedure. Majority of the HO failure instances are observed due to late HO attempts.
- Ping-pong HO KPI point out the early HO occasions in a cell. UE undergoing ping-pong HOs leads to back-and-forth HOs between the participating cells and can lead to higher signaling load and sometimes even low retainability KPI.
- Cell discovery KPI measure the small cell camping rate each time a UE is configured with a cell search process. Timely cell discovery can result in more offloading opportunities, and hence, efficient utilization of the available resources.
- Latency or data interruption KPI represents the delay UE observe during HO execution, paging attempt to success duration, accessibility etc.
- Accessibility KPI for a given time interval represents the percentage of idle mode UEs that were able to successfully acquire network access. Accessibility KPI indirectly impacts latency and user tracking KPI under rare circumstances for mobile users.

In most cases the KPI-parameter dependency is multi-pronged and leads to complex and often conflicting interplay between the KPIs and parameters. This interplay in the mobility KPI and the associated key parameters is summarized in Fig 8. The key challenges that arise from the convolved association between the mobility KPI and parameter [17] [18] are briefly described below:

1: Smaller qHyst value accelerates reselection, as soon as the target cell RSRP becomes greater than serving cell RSRP. As a result, accessibility issues related to idle mode mobility (as discussed earlier in the section) can be addressed. However, too low of a qHyst can result in unnecessary reselection (for instance, to an over-shooting cell).

2: Shorter Treselection will improve the accessibility KPI at the cell boundary due to timely reselection. However, too short Treselection will result in ping-pong reselection especially for stationary users (i.e. due to shadowing).

3: Idle mode Cell Individual Offset (CIO) to accelerate or decelerate reselection towards a neighboring cell. (configuring a positive CIO towards a particular neighbor can accelerate reselection, and vice versa)

4: Time window to evaluate mobility State [17] of a UE. Number of reselections made within this time window will dictate mobility state (normal, medium or high) of a UE. Reselection criteria is typically eased as mobility state changes from normal to medium or high.

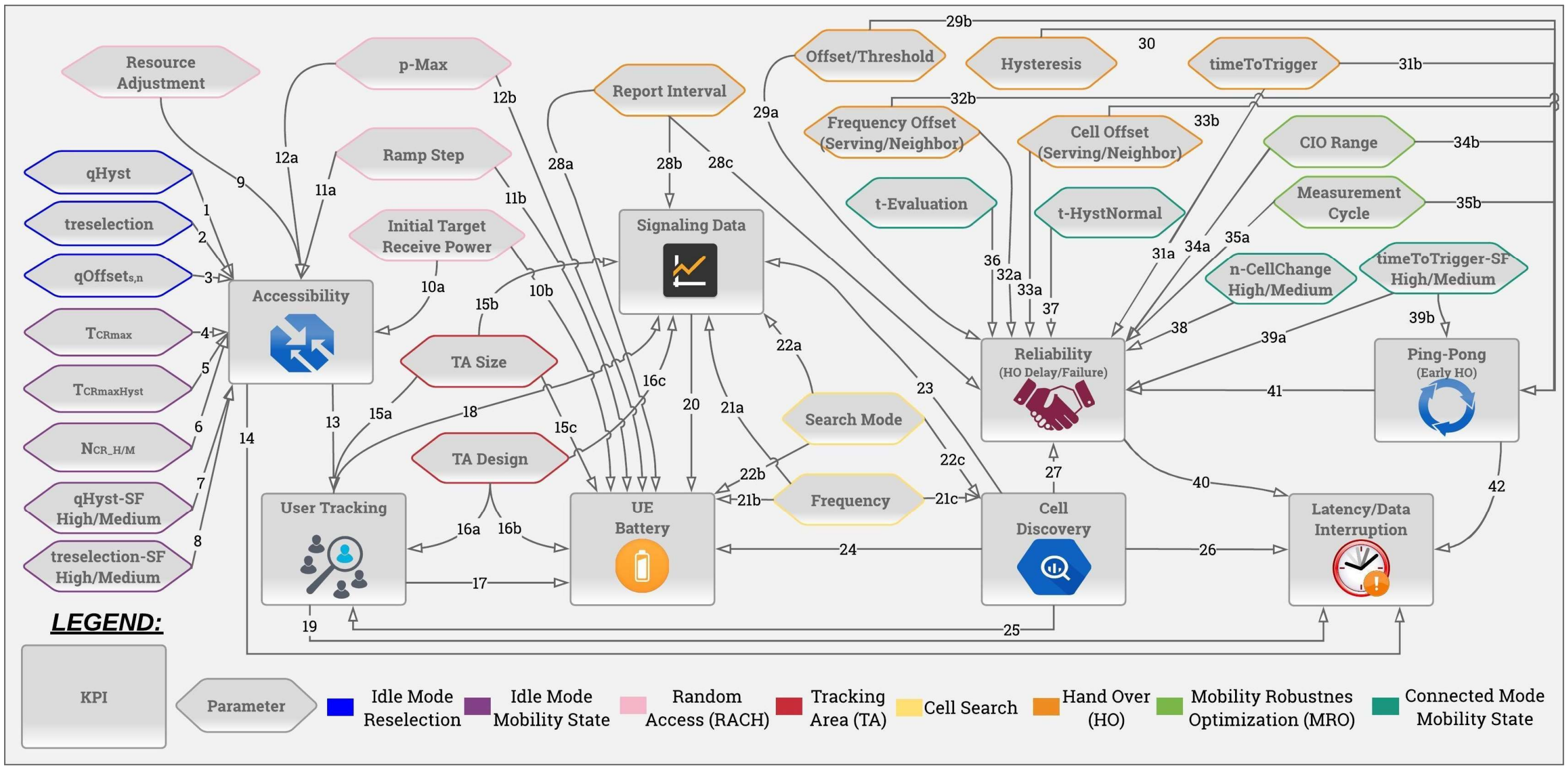

FIGURE 8. **Relationship diagram for mobility related KPIs and their interplay with the associated network parameters (grouped in different colors)**

**Source:** [17] [18]

5: Specify additional time period before UE can enter back to its normal mobility state with default reselection parameters. Recurrent mobility state change can be avoided by tuning this parameter.
6: Number of cell change needed (ignoring similar cells) within 'parameter #4' before UE changes mobility state from normal to medium or high respectively.
7: Scaling factor by which the default qHyst (parameter #1) is decreased when the mobility state is changed to medium or high.
8: Scaling factor by which the default treselection (parameter #2) is decreased when the mobility state is changed to medium or high.
9: Amount and location of RACH resources to ensure RACH success (providing adequate RACH resources, and avoiding RACH resource conflict between neighboring cells).
10: Higher target power can increase chances of RACH success at first attempt (better accessibility KPI) at the cost of a) higher battery consumption and b) chances of increased uplink interference for neighboring cells. An optimal target receive power is vital for better network operations.
11: Increase in the transmission power every time a RACH attempt fails. Higher step size can increase RACH success but with more battery consumption and vice versa.
12: Maximum allowable UE RACH power - Increasing maximum allowable UE transmission improves RACH success probability but with high energy consumption.
13: Improved accessibility to achieve a faster TAU can ensure accurate user tracking and prevent paging failure instances for mobile users.
14: Reduce latency through faster accessibility for mobile users (e.g. fast reselection to best signal cell and appropriate power for RACH success).
15: Smaller TA size will improve UE location estimate and will decrease the core network signaling due to smaller paging area. However, frequent TAU by mobile users will add radio access side signaling.
16: Suitable TA design (horizontal/vertical assignment) based on coverage conditions and type of traffic (e.g. high speed UEs) to ensure accurate user tracking and minimize TAU and hence, conserve UE battery and network signaling load.
17: Reducing TAU attempts for mobile users to conserve UE battery.
18: Reducing TAU attempts for mobile users to lessen signaling load.
19: Fast and efficient user tracking to reduce latency in accessing the network.
20: Minimizing signaling helps avoid unnecessary transmission and the UE battery can be conserved.
21: Higher cell search frequency will be beneficial to offload users to other cells. However, more battery will be consumed while searching. In addition, signaling load will increase every time a UE is configured with cell search procedure.
22: Periodic search mode will reduce signaling data generation as search configuration will be transferred to UE just once. However, small periodicity will waste the UE battery, and a large periodicity might miss a suitable offloading opportunity. On the contrary, a smart aperiodic search mode (e.g. location triggered) will be efficient and will save battery but signaling will be generated with each search configuration.
23: Signaling data generated for cell discovery purposes should be minimized.

24: UE consumes battery during cell search, hence, cell discovery should be minimized with high hit rate.
25: Timely cell discovery (intra-frequency) will prevent out-of-service (unreachable UE) occasions and Radio Link Failure (RLF) can be prevented.
26: Timely cell discovery (intra-frequency) will prevent recurrent re-transmissions and ultimately lead to Radio Link Failure at the cell edge.
27: Timely cell discovery (intra-frequency) will ensure HO success especially for mmWaves and the UE will not observe Radio Link Failure.
28: Smaller report interval (HO requests) will have more signaling data and battery utilization. However, the reliability KPI will improve as there will be more chances of BS being able to successfully receive and decode the HO request.
29: HO offset/threshold can be tuned to achieve timely HO.
30: Suitable hysteresis parameter will minimize chances of ping-pong HOs.
31: Small timeToTrigger can result in ping-pong HOs (e.g. for non-mobile users), while long timeToTrigger can avoid the HO resulting in low reliability/retainability KPI (e.g. to overshooting cells). Similarly, high speed users should be configured with lower timeToTrigger to accelerate HO to cell with best RSRP.
32: Frequency based CIO to accelerate or decelerate inter-frequency HOs to all neighboring cell(s). Optimal CIO can prevent late and/or early HO.
33: Relation based CIO to accelerate or decelerate intra/inter-frequency HOs toward the configured neighboring cell(s). Optimal CIO can prevent late and/or early HO.
34: Configuring a large CIO range can avoid the chances MRO assigns a large CIO (a large CIO is not recommended as it can have negative consequences especially for static users)
35: Shorter MRO cycle can recommend suitable CIO configuration based on changing traffic conditions. However, too short of a cycle should be prevented as it can have sub-optimal recommendations due to inadequate statistical data required to configure optimal CIO.
36: Similar to 'parameter #4' but for connected mode.
37: Similar to 'parameter #5' but for connected mode.
38: Similar to 'parameter #6' but for connected mode.
39: Similar to 'parameter #8' but for connected mode.
40: HO failure results in higher latency and more data interruption occasions.
41: Frequent HOs increases the risk of HO failure both for static and mobile users.
42: Latency and data interruption are intrinsic to break-before-make HOs, hence ping-pong HOs should be avoided.

Fig 8 illustrates the simplest representation of the complex interaction between various KPIs and mobility related network parameters. It can act as a foundation, with the help of which, researchers can devise an ideal mobility management scheme that aims to minimize the negative impact on KPIs indirectly affected by tuning mobility related network parameters. Now, we present a detailed survey of the state-of-the-art literature available on mobility challenges and corresponding research proposals. *Insights from this tutorial section will be leveraged to evaluate the research papers in terms of conflicting KPI(s).*

## III. MOBILITY CHALLENGES AND RESEARCH PROPOSALS

Seamless mobility experience at a very high-speed is considered as one of the major use cases for 5G networks, particularly in wake of advent of autonomous cars, low altitude drones, and emerging high-speed commute systems. The mobility characteristics of the emerging networks, such as densification and adaptation of mmWave narrow beam cells (discussed in section I), combined with the intrinsic complexity of the mobility management process (discussed in section II) means that the mobility management in 5G and beyond requires significant research efforts by wider community. In this section, we review the recent contributions made by the research community to address 5G and beyond mobility challenges, by categorizing them in six sections as shown earlier in Fig. 1. Studies focused on reliability goals that involve achieving seamless and timely HO while preventing HO failures and ping-pong HOs are discussed in the first sub-section. Studies focused on achieving mobility while maintaining small delay are discussed in the Latency Requirements sub-section. Signaling Minimization approaches are presented in the next sub-section, followed by User Tracking in futuristic ultra-dense networks. Subsequent sub-section covers studies on cell discovery including the goal to perform timely offloading from macro-cells to small-cells in order to prevent network congestion and efficiently utilize network resources. Finally, research work focused on lessening energy consumption are presented in the last sub-section.

### *A. RELIABILITY GOALS*

Mobility casts a serious threat to reliability especially when HO is being performed from one cell to another. Now we will discuss different research work on different HO types and the respective reliability goals. Comparison of reliability enhancement approaches has been presented in Table VI.

#### 1) BREAK-BEFORE-MAKE AND RELIABILITY

5G NR employs break-before-make (hard) HO approach [18] where UE breaks the connection with the serving BS before resuming the new connection with the target BS, and this process makes the mobile UE prone to undesirable service interruption. Repetition of this type of HO under ping-pong scenario makes it even more susceptible to call drops. An effort to deal with the frequent HO case has been presented in [19]. This paper focuses on the multi-objective learning-based mobility management strategy where a learning model is described to obtain a comprehensive network information. Then a multi-objective mobility management method is proposed taking into consideration user QoE and number of HOs. Results are compared with 3GPP based HO scheme, and the authors show that number of HOs are reduced by more than 5 times. As a future step, simulations can be presented by using a stochastic network model.

TABLE VI
RELIABILITY ENHANCEMENT APPROACHES

| References | Pros | | | | | | Cons | | |
|---|---|---|---|---|---|---|---|---|---|
| | **RLF Prevention** | **QoE Improvement** | **Uplink Consideration** | **BER Reduction** | **#HO Reduction** | **Latency Reduction** | **User Velocity Not Considered** | **Throughput Degradation** | **High Complexity** |
| [19] | | ✓ | | | ✓ | | | | |
| [20] | | | ✓ | ✓ | | | ✓ | | |
| [21] | ✓ | | | | | | | ✓ | ✓ |
| [23] | ✓ | | | | | | ✓ | | |
| [24] | ✓ | | | | | | ✓ | | |
| [25] | ✓ | | | | | ✓ | | | |
| [26] | ✓ | | | | | | | | ✓ |
| [27] | | | | | | | | ✓ | ✓ |

Much of the reliability concerns are studied while keeping in view the UE downlink performance only. Authors in [20] studied reliability for uplink channel of multi-user MIMO channel. Authors employed Quadrature Spatial Modulation (QSM) to lower the uplink Bit Error Rate (BER) from $10^{-1}$ (when using spatial multiplex) to the order of $10^{-3}$. As a future work, BER results can be shown with different user velocity to evaluate the efficacy of the proposed approach for a realistic scenario of mobile users.

#### 2) MAKE-BEFORE-BREAK AND RELIABILITY

Unlike 5G NR and LTE, 3G uses an alternative of break-before-make HO, i.e. make-before-break vis-a-vis soft HO. 3G UE apply macro diversity where it can establish simultaneous connection to more than one cell, and the set of participating cells are referred to as Active Set (AS). Authors in [21] propose a 3G like soft HO approach where multiple serving cells are represented by AS. The results show that fixed AS window can prevent RLF to a great extent. However, throughput degradation is observed as radio resources of the weaker cells are unnecessarily wasted by the user. To counter this problem, the authors propose a dynamic AS window where add/remove parameters are adapted based on the slope of the linear curve that creates the dependency between the add/remove offset and the size of AS. AS based approach will result in more signaling, computation and energy requirements in maintaining and updating the connectivity to different cells in the AS.

One drawback of make-before-break HO scheme is the complexity at UE side to process multiple RF chains. Note that the advent of narrow mmWave beams in 5G that is likely to lower the source link reliability for the mobile users, further undermines the perceived advantages of make-before-break HO. Authors in [22] analyzed the pros and cons of make-before-break HO in more detail and concluded that they are unsuitable for 5G networks. For similar reasons, 3GPP RAN WG2 during its meeting #94 decided to discard make-before-break like procedures from the scope.

For the above-mentioned reasons and to achieve higher reliability and retainability goals, the 5G networks have employed hard HO process requiring successful break-before-make procedures. Reliability goals in literature are usually addressed through multi-connectivity approaches.

#### 3) RELIABILITY THROUGH MULTI-CONNECTIVITY

Multi-Connectivity (MC) can be employed in conjunction with break-before-make HO approach to mitigate interference through coordination. MC can attain ultra-reliability, low latency, and interruption-free communication by preparing the target cell before the transmission is broken. Furthermore, it tackles connection failures by using a coordinated transmission among the serving cells. As a result, HO failures and RLFs are drastically suppressed. However, drawback of MC includes added complexity in adding/removing MC participant cells. A study by Tesema et al. [23] on intra-frequency MC shows that the RLFs can be avoided while enhancing throughput through joint transmission of BSs. The authors in [23] then extended their idea in [24] to inter-frequency MC and prove availability benefits in that scenario. However, stationary users were considered with focus on modeling of the best server association. Their study did not incorporate reliability for mobile users.

In a separate study [25], the same group of authors deal with mobility concerns and evaluated reliability performance through different intra/inter frequency cells. For intra frequency, Dynamic Single Frequency Network (DSFN) is proposed to dynamically add BSs to the coordination set. This in turn helps to achieve reliability and low latency of less than 1ms. For inter-frequency on the other hand, redundant transmissions are performed on the different frequency layers, such that the UE selects the best transmission, i.e., selection combining is applied. The proposed approach can avoid poor SINR of <-6dB (marked as RLF) and achieve higher reliability of 99.999% or greater.

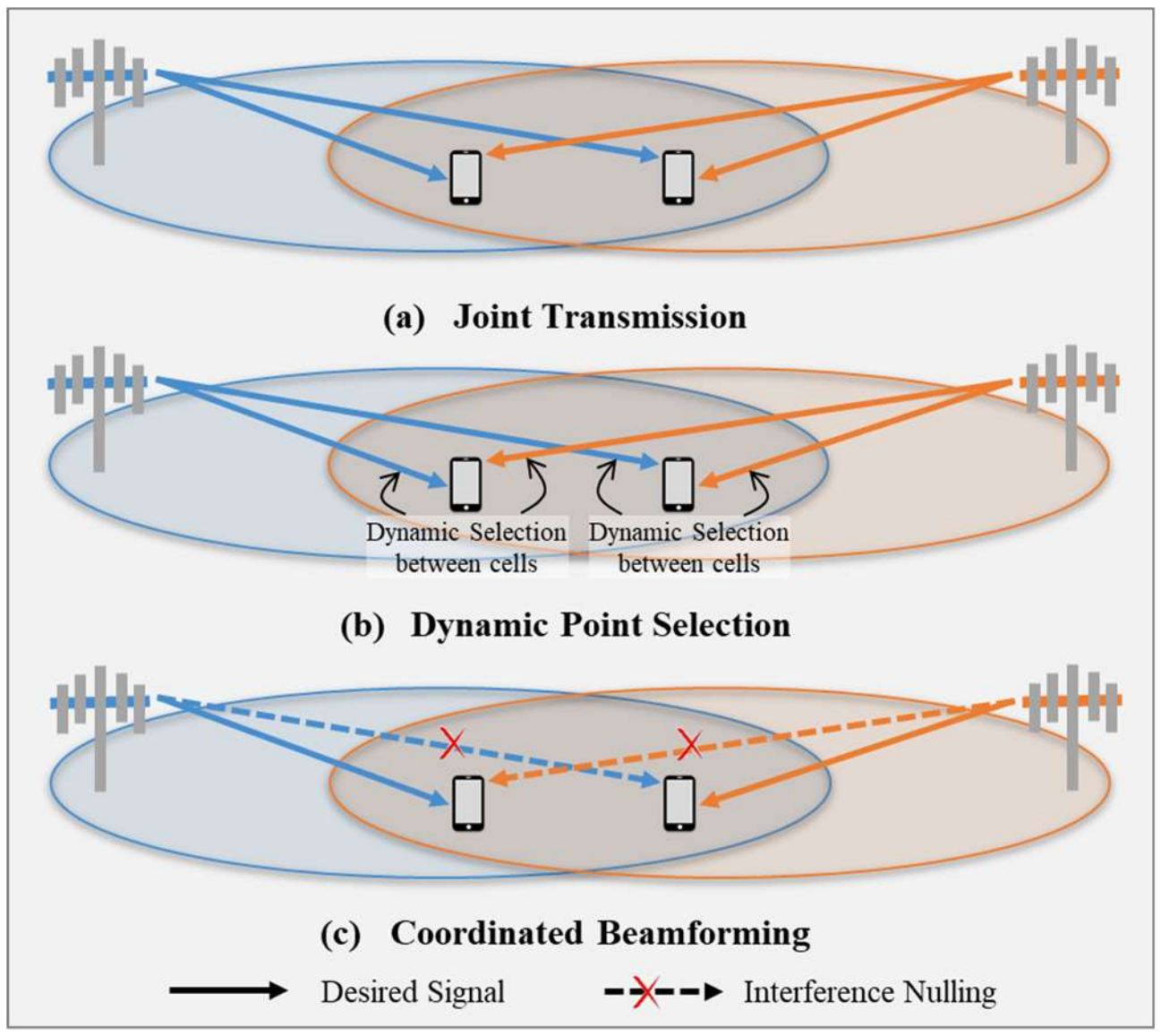

FIGURE 9. **Types of Downlink CoMP.**

Tesema et al. further enhanced their work in [26] by proposing a novel multi-connectivity scheme that uses fast selection of serving cell from a set of prepared cells similar to Co-ordinated Multi-Point Transmission (CoMP). Fig. 9 shows different types of CoMP. Control plane in CoMP is served by a primary cell only, and if radio condition of the respective control channel degrades, then user plane data may not be guaranteed even if radio condition of user plane cell is better. On the contrary, Fast Cell Select (FCS) is proposed in which the selected cell from the set of pre-arranged cells is used for transmission of both data and control signals. The presented work provides gain in the quality of the control and data signals, which ultimately solves RLF problem and improve throughput of cell-edge user.

CoMP, although beneficial, has an intrinsic conflict with the hard-HO methods used in 5G networks, as connection with source cell terminates before setting up a connection to the target cell. In [27], authors addressed this conflict by introducing a new HO mechanism based on CoMP joint transmission scheme in order to minimize inter-cell-interference (ICI) level between the adjacent cells during the HO execution. Their algorithm consists of Coordination set (CS) and Transmission set (TS) of BSs. CS selection is assisted by the UE through sending periodic measurement report which contains UE velocity and RF condition. Velocity metric is used to avoid small-cells for high velocity UEs, and RF condition is used to determine TS. Performance evaluation results show that ICI is reduced considerably leading to a better average throughput per user during the HO procedure. Benefits are achieved at the cost of higher complexity and increase in signaling data. A study on optimal TS size to improve reliability, and throughput, taking into consideration the processing complexity and the magnitude of the control data would be a good research contribution.

### B. LATENCY REQUIREMENTS

Besides reliability, another mobility management objective of paramount importance is to minimize the length of transmission disruption during the HO process. In this subsection we review the studies and research efforts aimed to minimize HO delay.

#### 1) RACHLESS HANDOVER

Authors in [22] identified that RACH takes about 8.5ms out of 50ms interval required to accomplish HO task in LTE. Based on this assumption, they proposed a RACHless HO technique to improve the latency by 17%. Authors suggest alternate means to perform the same functionalities as of RACH. For instance, RACH helps target BS to compute Timing Advance, though with lower accuracy. In the proposed RACHless HO, UE can estimate timing advance from the time difference between the source and target cell signals. Accuracy evaluation of the proposed approach will help gain confidence to the researchers. Such timing advance estimation method has been further evaluated in [28]. Alternatively, target BS can also compute timing advance through Sounding Reference Signals (SRS) which is used in LTE for uplink channel estimation as shown in [29]. However, this process might result in the timing advance estimation delay as it requires UE to be configured with SRS first. Initial uplink power, Physical Uplink Shared Channel (PUSCH) in LTE, normally known after successful RACH procedure, can be determined through source BS prior to HO initiation. Eliminating RACH is a novel proposal. However, UE in turn has to do more processing to compute timing advance that may lead to decreased battery life in a dense network.

While RACHless HO has its merits, the aforementioned challenges call for alternative approaches to reduce HO latency. One example of such approach is mobility aware caching.

#### 2) MOBILITY AWARE CACHING

From the mobile users' perspective, more data rate alone is not enough to ensure better user experience. Any bottleneck in the distribution network between RAN and content servers can result in a prolonged Round-Trip-Time (RTT). During a HO, the chances of such bottleneck increase as momentarily the UE's QoE becomes dependent on two cells instead of one. This makes caching in the BS a useful tool to help accelerate the data delivery to the intended user. However, mobility degrades cache efficiency when UE moves to another BS. A study in [30] proposes to incorporate caching and computing ability deep into the base stations. The authors in [30] proposed a seamless RAN-cache HO framework based on mobility prediction algorithm (MPA). In the proposed scheme, the target BS is predicted for a UE with unfinished transmission during HO. This prediction is then used to pre-trigger the source RAN cache. This notifies the target RAN cache associated with the target BS to prepare for serving the UE and ultimately reducing latency. As a result, false probability of RAN-cache HO pre-trigger through MPA though recorded to be less than 1.36% show an 8% increase in the maximal RAN-cache HO processing time. Researchers should benefit from the history of user mobility to come up with an improved algorithm.

Mobility aware caching has been investigated in [31] to maximize the cache hit ratio that is defined as the number of requests delivered by the cache server, divided by the total number of requests. Compared to [30], authors in [31] considered both macro-cells and small-cells. The first priority is given to the local cache followed by small-cell. However, if data is not received within the set deadline, macro-cell is then accessed to acquire data. Results assert that the proposed caching strategy outperforms prior caching strategies. The proposed cache scheme has a better cache hit ratio and low latency requirement for 5G networks.

#### 3) PAGINGLESS APPROACH

Authors in [32] presented a novel frame structure with sub-millisecond subframe duration operating in Time Division Duplex (TDD) mode aimed for 5G networks. The frame structure carries UL beacon resources to enable a pagingless system for idle mode users. For connected mode users, UL beacons provide channel state information (CSI) for improved frequency selective scheduling. However, a caveat of this approach is that it can lead to an excessive amount of uplink messages. This in turn, may cause accelerated UE battery drainage and thus smaller battery life which is contradictory to one of the major 5G requirements.

### C. SIGNALING MINIMIZATION

In both LTE and 5G NR, the processing unit is shifted to the edge, i.e., BS, primarily to reduce latency. However, this comes at the expense of increased signaling generated as the UE context is shifted from one cell to another during the HO procedure. This issue aggravates with the ultra-dense BS deployment. High signaling not only chokes the CPU of BSs, but also results in lower effective spectrum efficiency by consuming a substantial amount of resources in the air interface. Too much signaling between neighboring BSs and BS-Core can result in potential congestion in the backhaul for the 5G networks with ultra-dense BS deployment. Reason being the expected myriad of mobile UEs, ultra-dense BS deployment, and added features that require high coordination e.g. multi-connectivity, carrier aggregation, and interference mitigation techniques. Thus, there is a possibility of network being paralyzed especially in busy hours due to the avalanche of signaling traffic. Signaling avalanche is an eminent threat in future ultra-dense networks. The research efforts by the research community to minimize the mobility signaling load can be loosely categorized in the following four sub-categories.

#### 1) HO SIGNALING REDUCTION THROUGH MINING HO PATTERNS

One basic but effective way to reduce HO signaling is to characterize HO behavior among cells to identify cells with an unusually large number of HOs or otherwise abnormal HO pattern e.g. ping-pong. Authors in [33] study the HO behavior of cells and propose a clustering model using K-means, to group cells with similar HO behavior. Further evaluation was done using actual HO attempt and HO success KPI of nearly two thousand WCDMA cells. The idea is to forecast the number of HOs and detect abnormal HO behavior among cell pairs using linear regression and neural network techniques. The detection is then used to perform targeted optimization of HO parameters in respective cells to minimize HO signaling. Adding a temporal component to training data can further increase the accuracy of the prediction.

#### 2) MOBILITY SIGNALING REDUCTION THROUGH RAN CENTRALIZATION

Another method to reduce mobility signaling is to leverage the centralization of RAN e.g. using Cloud-RAN (C-RAN). Uladzamir et al. [34] recently proposed mobility aware hierarchical clustering approach (HIER) to group Virtual Base Stations (VBSs). Clustering based on the location of Radio Resource Heads (RRH) aims to reduce costly HOs and thus, minimize signaling data. They also proposed location aware packing algorithm (LA) where inter-cluster mobility statistics are obtained by keeping track of UE movement, UE history to predict the traffic intensity between BSs. In addition, the history of inter-RRH HOs is considered as well. The proposed scheme when compared with affinity propagation clustering [35] can reduce up to 34.8% HOs, but at the cost of much higher requirement of RRHs. The approach can be beneficial for urban areas, but for less dense sub urban and rural areas, network deployment at this scale won’t be feasible.

#### 3) MOBILITY SIGNALING REDUCTION THROUGH CELL EXTENSION

An Extended Cell (EC) concept is proposed in [36] to dynamically form groups of several adjacent cells. HO performance improvement is rendered by increasing the overlapping area between two adjacent cells in the Radio over Fiber (RoF) indoor networks. The proposed approach reduces the number of HOs and the call drop probability during the HO by 70%. Although proven effective, it lacks the dynamic procedures to define ECs to optimize network resources. Shortcomings were addressed by authors in [37] by extending the idea and coming up with a proposal on the Moving Extended Cell (MEC). Here, each mobile UE is covered by 7-cell EC where each EC transmits the same user data at every instance. This in turn, reduces HO latency through early preparation. Evaluation results show the proposed architecture can totally avoid call drop and packet loss for UE’s with a velocity of up to 40 m/s. The authors in [37] suggested that MEC is very efficient in tackling HO for mmWave cells but is vulnerable to throughput inefficiency as all seven cells in the cluster transmit for a single user.

#### 4) MOBILITY SIGNALING REDUCTION THROUGH VIRTUALIZATION

Virtual Cell (VC) has been proposed as a solution by Hossain et al. in [38] to reduce mobility signaling while increasing the throughput efficiency of 60 GHz RoF network. VC is a central part of an actual cell, and the remaining boundary area is divided into numbered tiles. Wireless Sensor Network keeps track of the UE location and periodically sends report to a centralized controller. Multiple Antenna Terminals (AT) cover a single cell, and only a single AT is activated at an instant. When the UE steps on one of the boundary-located tiles, the controller activates respective neighbor AT to transmit similar data. In the VC scheme proposed in [38], maximum of only two ATs can be activated for HO preparation in contrast to 6 in MEC [37]. End results of using VC concept show an increase of 33% throughput efficiency in comparison to MEC. Drawback of the proposal involves management of a wireless sensor network to track and report UE location. And if the UE velocity is high, the low powered sensors may not be able to timely report or even identify the presence of a high-speed user.

### *D. USER TRACKING*

Location management, sometimes referred to as mobility tracking or user tracking, is defined as the set of procedures that determines UE location at any instance. User tracking is inevitable in cellular networks, so that incoming data from the core network can be delivered to the user. Densification of both cells and users, as well as increased mobility focused use cases such as Intelligent Transportation Systems (ITS)/Unmanned Aerial Vehicles (UAV) etc. bring new challenges to user tracking in 5G environment. The recent attempts to address these challenges can be loosely categorized into following three subcategories:

#### 1) DISTRIBUTED TRACKING AREA UPDATE

A framework to minimize conflicting metrics, Tracking Area Update (TAU) and paging, is presented in [39] by distribution of Tracking Area (TA) into Tracking Area Lists (TAL) in two phases. First phase is offline, which is responsible to assign TAs to TALs using three different approaches. The first two favors paging overhead and TAU respectively, while the third one uses nash bargaining game to ensure fairness between paging overhead and TAU. Second phase is online which controls the probabilistic distribution of TALs on UEs by taking into account their behavior, incoming transmission frequency and mobility patterns. Numerical results were shown for the three approaches of the first phase, where the third solution provides a fair tradeoff between paging overhead and TAU. As a future step, results should be compared with prior schemes.

No research work focusing on the horizontal or vertical deployment of TAs is present, therefore researchers can come up with smarter and more effective ways for operators to define Tracking Areas.

#### 2) HYBRID TRACKING AREA UPDATE AND PAGING

5G network will have large range of UEs and dense network deployment as discussed earlier. Hence, a huge amount of paging especially for millions of IoT devices is expected. As a result, signaling associated with paging may become enormous if currently available approach is used. To address this problem, authors in [40] propose a hybrid scheme in which either RAN or core network can initiate paging. RAN based paging with Tracking Area (TA) of just one BS is proposed for the RRC inactive [41] UEs to have low latency at the expense of high buffering capacity to transfer the content to the neighboring BS in case of user mobility. Meanwhile, core network-based paging is recommended to be used for idle UEs. Authors also proposed a hierarchical paging and location tracking scheme to minimize signaling load by assigning an anchor BS for location management. They conclude that RAN based paging is not efficient for high mobility UEs as TA is limited to a single BS. For hierarchical approach on the other hand, there should be more data management and processing for every user at anchor BS which becomes another single point of failure. Processor overload or X2 (inter-cell communication link in LTE) congestion, as a result, can disrupt the paging process.

#### 3) DYNAMIC/ADAPTIVE TRACKING AREA UPDATE

Authors in [42] proposed an adaptive method that employs smart TAs to reduce the frequencies of TAUs and the sizes of paging areas. The proposed scheme uses the interacting multiple model (IMM) algorithm [43] to determine the estimated location of a UE at the time of the latest registration and provide a predicted location after a certain time frame. An experimental evaluation with an artificial trajectory showed that this approach cuts half of the extra location registrations compared with non-adaptive methods. Aside from that, this method also determines TA adaptively to significantly reduce the average paging sizes resulting in to lesser signaling for each paging attempts. As a future step, comparison results can be added for different types of mobile users at different speeds and trajectories to prove the effectiveness of their approach.

Authors in [44] employed Apriori algorithm [45] for dynamic Location Area planning using call logs of several mobile users. Apriori algorithm finds frequent itemset using an iterative level-wise search procedure. By taking minimum support of 100%, Apriori algorithm can highlight those cells which serve mobile users every day. Based on this approach, authors in [44] suggested to create a dynamic TA based on more than 80% minimum support. Authors in [44] categorized mobile users into predictable, expected and random groups based on the minimum support value. For each category, the authors propose to minimize location management cost by employing a suitable algorithm. However, the exact algorithms needed to minimize location updates, in this scheme, remain to be investigated as future work.

### E. CELL DISCOVERY

Traditional networks with High Frequency (HF) bands broadcast the reference signals (pilot symbols) for cell discovery as mandated by 3GPP. Majority solutions proposed in literature for cell discovery involve periodic scanning by the UE of these broadcast signals. The higher frequency of this periodic scanning ensures timely cell discovery but results in increased battery consumption leading to trade-off between energy efficiency on UE side, network side, QoE, overall capacity and load distribution. In the following we discuss studies that have investigated these trade-offs and proposal solution to optimize one KPI or other.

#### 1) CELL DISCOVERY WITH UE ENERGY CONSTRAINT

5G networks will have heterogeneity of BSs with a motely of macro-cells and small-cells. A mobile UE connected to a macro-cell must scan for potential small-cells to benefit from the high data rate and traffic offloading opportunity. If a mobile UE uses high scanning periodicity, it is likely to discover small-cells in a more timely fashion. Thus, it may avail better offloading opportunities, but at the cost of reduced battery life due to increased amount of energy consumed by the scanning process, and vice versa. The investigation of this tradeoff is interesting and yet a challenging research problem as the optimal scanning periodicity, if exists, might be dependent on the cell density and user speed among several other factors.

Authors in [46] use a rigorous approach that leverages stochastic geometry-based modelling of the network and empirical modeling of UE mobility. Analytical expressions have been derived to characterize and quantify the dependency of the UE energy efficiency on the cell density, cell discovery periodicity and the user velocity. Through analytical as well as Monte Carlo simulation results, it's been shown in [46] that UE battery life reduces significantly with increased cell discovery rate, while the UE throughput increases and vice versa. The key finding of this analysis is that, there exists an optimal cell discovery frequency for a given cell density and user speed statistics. This optimal cell discovery frequency maximizes the UE energy efficiency (EE) by achieving a Pareto optimal point between the capacity lost by missing cells with low cell discovery frequency and energy saved at UE in doing so and vice versa.

Daniel et al. [47] proposed an energy efficient small-cell discovery technique using radio fingerprints. In this proposed solution, network configures UE with several radio fingerprints which are lists of cell-IDs and RSRP strength at different intervals. As a normal procedure, users served by the macro-cell performs the neighbor cell measurement as it moves around and compares those to the configured radio fingerprints. Upon a successful match, macro-cell is reported back which in return configures the corresponding small-cell. Authors show that energy efficiency of 70-80% is achieved on UE side by avoiding unnecessary small-cell discovery measurements, and up to 45% on network side by small-cell activation/deactivation. Practical use of this approach will be limited to shadowing since RSRP at a given point changes with time and the effect of environmental changes like rain/snow also affects the standard deviation of shadowing. Moreover, MDT will reveal better results as the location of the UE with respect to the small-cell location can be known, followed by the successful small-cell association.

#### 2) CELL SELECTION WITH NETWORK ENERGY EFFICIENCY PERSPECTIVE

The Information and Communications Technology (ICT) sector contributes around 2-3% to world's carbon emissions and is doubling every four years [48]. Since mobility is closely coupled with uneven and dynamic user distribution, the mobility patterns can be exploited to turn OFF/ON cells for enhancing energy efficiency. A solution to conserve network energy using such mobility leveraging approach is proposed in [48]. Decision of powering OFF the BSs is made using the UE velocity, receive power, BS load and energy consumption. In addition, HO to the small-cell can be made only if the UE velocity and the cell load is lower than the respective thresholds. As a result, the low load cells can be powered OFF. However, the paper does not address when and how to turn ON the cell, as the powered OFF cell in the presence of the candidate UEs can have negative impacts on the capacity, efficiency and user satisfaction.

Random way point mobility models and the stochastic geometry theory are utilized in [49] to evaluate the energy efficiency of 5G networks. The network capacity and energy efficiency are evaluated for Ultra-Dense Cellular Networks (UDN) considering the user mobility. Results were demonstrated using Monte Carlo scheme where a user will keep stationary for a certain time, and then start moving to a random direction with variable but bounded velocity range. Results indicate that the energy efficiency decreases exponentially with increase in the small-cell density. Energy efficiency decreases from 160bits/J to 155bits/J and 144bits/J when small-cell density was increased from 10 cell/km$^2$ to 15 cell/km$^2$ and 20 cell/km$^2$ respectively.

#### 3) MMWAVE BEAM ALIGNMENT AND USER TRACKING

The studies discussed in the last two subsections do not consider the several idiosyncrasies arising from the advent of mmWaves cells, as discussed in the following. mmWave band cell discovery becomes far more complex compared to the high frequency (HF) cells because of the high penetration loss and narrow beams [50].

Directional path in mmWave can deteriorate sharply due to rapid changes in the environment which calls for an intense tracking and alignment. The situation can be aggravated when considering mobile users. To address these issues, authors in [51] proposed two innovative schemes by which UE can alternately scan the whole angular space exhaustively and select the beam with the best SINR. They propose the mmWave BS to send pilots in the configured finite directions at regular intervals, one at a time. The UE then scans for the mmWave-cell beam using two mechanisms: a) periodic refresh (PR) – The UE scans in all directions one at a time and the direction with the maximum SINR is selected; b) periodic refinement and refresh (PRaR) – The first optimal beam with the maximum SINR is selected as per the PR, and then the UE performs a refinement procedure by scanning the neighboring direction to adapt according to the changing condition or due to the UE mobility. This mmWave tracking approach is depicted in Fig. 10. Comparison between both schemes were done using the real-world measurement data collected in New York city on carrier frequency of 28GHz. As expected, PRaR is less energy efficient than PR because of the much frequent refinement procedure. However, they did not compare their schemes with the broadcasting approach or direct alignment schemes. Also, the scenario might arise where both the mmWave BS (in sending pilots) and the UE (in scanning pilots) are not synchronized with each other in terms of direction. Such a scenario is likely to lead to the tracking and alignment delay. Alignment process is done by scanning the adjacent beams only and can give sub-optimal results for the high-speed users.

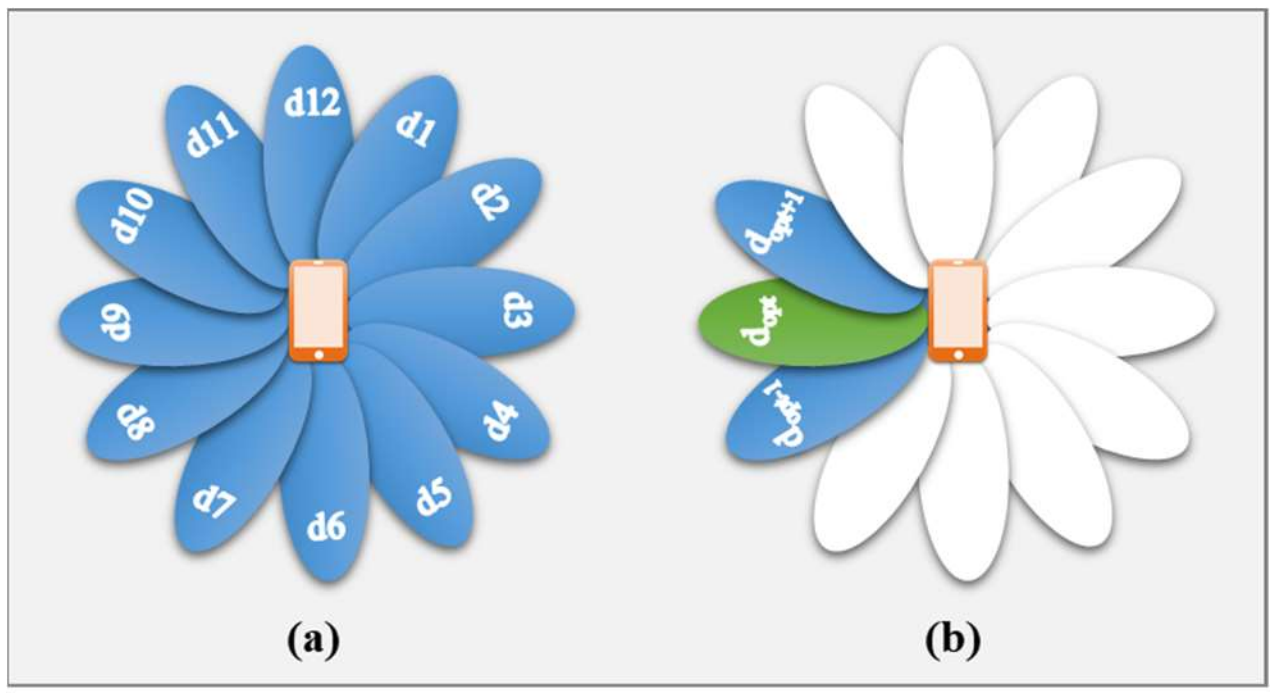

FIGURE 10. **mmWave tracking. (a) Refresh procedure through 12 directions, (b) Refinement procedure through 2 directions.**

Esmaiel et al. [52] proposed a novel mmWave multi-level beamforming approach. mmWave link is established after multi-level beam search is conducted using a compressive sensing-based channel estimation. The estimated UE location is used to determine the number of beams and the bandwidth required for constructing the sensing matrix used in each beam searching level. Results show an increase in the spectral efficiency by 40% under good radio conditions. Authors in [52] also proposed a novel concept [53] of two-level control and user data (2CU/U) planes splitting, where the LTE BS and the WiFi access point provides control over the distributed sub-clouds and distributed mmWave BSs respectively. With the proposed approach, mmWave miss-detection probability as low as 10% can be obtained compared to 90% with the conventional approach when mmWave BS are deployed in a sparse manner. The result can be further improved by incorporating the user movement historical data, and to observe the result for different UE speed.

#### a) *HO in mmWave Band*

Traditional HO is based on the Received Signal Strength (RSS) wherein pilot signal strength measured by the UE determines the cell-edge and thus lends assistance in performing HO to the target cell. This approach is ineffective for addressing the unique challenges associated with the mmWaves. In mmWave cells, the RF reception changes drastically with UE speed and direction. Hence relying on the RSS to anticipate a cell edge may not suffice.

Authors in [54] suggest a novel Inter-Beam HO Class (IBHC) concept combined with the HO control and radio resource management functionalities. Initially, the user is assigned to a mobility classes depending on its estimate speed. The corresponding HO frequency is defined such that the high velocity UEs are expected to observe more HOs than the pedestrians. The mobile user is assigned a group of beams as per mobility class, load conditions and the expected path of UE. Each beam in the group contains similar resource allocation to improve the reception quality. HO is thus performed only at the edge of the beam-group. The underlying assumption in the proposed scheme is that the individual signals of each beam are perfectly synchronized. This can be true for low speed users; however, it may not hold for the high-speed users. Another strong assumption is the perfect estimation of UE velocity. UE velocity estimation is a big challenge even in the existing mobile networks, where the number of HOs in a moving time window are used to estimate UE velocity. Emerging networks with dense deployment of multi-frequency networks will make the prediction of UE velocity even a bigger challenge. Concept presented in the [54] can be extended by considering the relationship between the maximum user velocity and the mmWave footprint where its beneficial for the mobile user to camp to the mmWave cell. The study should include the signaling cost and energy consumption in scanning for the mmWave cells.

In [55], authors leverage the concept of moving cell for train communication using 60 GHz band. To avoid the large number of HOs in high speed train, authors propose to employ the Radio over Fiber (RoF) technique. The key idea is to make the serving cells move together with the train and thus provide smooth uninterrupted transmission to the passengers. However, for this scheme to be practical, the train's velocity and the direction needs to be pre-known to achieve synchronization. Furthermore, due to the inability to cope up with randomness of user mobility, this concept is not appropriate for mobility management in indoor environments.

The state-of-the-art literature work reviewed in this section is focused on managing mobility in a reactive way. Two of the key challenges in mobility management in emerging networks that are not addressed by the current reactive mobility management paradigm in the industry and the associated literature in academia are high latency of the HO process and the large signaling overhead. These challenges become more important with the increasing fraction of mobile UEs, more bandwidth hungry applications and the advent of delay sensitive use-cases like self-driven vehicles. Proactive mobility management is an emerging paradigm that has the potential to address these challenges. It's a vital component by which the network operators can guarantee the success of the futuristic mobile networks. Key concept of the proactive mobility management and the recent studies that have presented few novel ideas to achieve the proactive mobility management are discussed in the next section.

## IV. PROACTIVE MOBILITY MANAGEMENT

It is a well-researched fact that people tend to visit the same places repeatedly in their daily life, e.g. workplace, school, gym, parks, shopping venues, etc. This makes their movement to feature a high degree of repetition and hence predictability. According to some large-scale studies, this perceptibility can be as high as 93% [56]. This intrinsic predictability in human mobility can be leveraged to build models to predict the UE mobility patterns. In cellular networks, these models can be built by harnessing the large volumes of UE mobility related data such as call detail records (CDRs), GPS traces, and data traffic from existing networks. Following is the list of some of the potential use cases of mobility prediction in the current and emerging cellular networks:

- Enhancing the overall QoS and QoE by reserving and managing radio resources a priori for users expected to arrive in a cell [57].
- Prevent failures and minimize HO delay e.g. by proactively triggering HO [58] [59].
- Prevent ping-pong HOs.
- Efficient load balancing e.g. by predicting cell loads and emergence of hot spots.
- Assist in cell activation/deactivation, and hence, conserve energy consumption.

Mobility prediction models in literature can be classified into three broad groups:

1) History based prediction models: In this type of prediction models, UEs next target cell is predicted based on the statistical analysis of historical records such as HO records or CDR records.
2) Measurement based prediction models: Such prediction schemes derive probability of user transition to next cell based on the real time measurements e.g. RSSI, SINR, distance, etc.
3) Location based prediction models: Current user location and in some cases urban transportation infrastructure is used to predict the future user location in the location-based prediction models.

In the following, we discuss the recent studies in literature that have made use of the two types of prediction approaches for various use cases.

### *A. History Based Prediction*

History based mobility prediction approaches can be further divided into the following categories:

#### 1) CELL TRACE BASED PREDICTION

Location prediction based on cellular network traces has recently attracted a lot of attention. Zhang et al. propose NextCell scheme [60] that utilizes social interplay factor to enhance mobility prediction. Social interplay is characterized by the convolution between entropy of the average call duration between two users, and the probability distribution of these two users to be co-located in the same cell. NextCell predicts the user location at cell tower level in the forthcoming one to six hours. It shows that inclusion of the social interplay improves prediction accuracy by 20% when compared to behavior periodicity-based predictor. However, results were not compared with the existing prediction schemes.

Authors in [61] presented a HO prediction scheme that combines signal strength/quality to physical proximity along with the UE context in terms of speed, direction, and HO history. The presented scheme achieves 33.6% reduction in HO latency when compared with conventional HO approach.

#### 2) MACHINE LEARNING BASED PREDICTION

Complex interaction between different components of a network can be well captured by Machine Learning approaches. For the same reason, much of the history-based prediction works revolve around machine learning based approaches. Authors in [33] argue that most of the research involving behavior prediction of a single UE is an infeasible and impractical approach. The argument is backed by the fact that some HOs are coverage based, while some are network initiated (e.g. load balancing). They propose to address these challenges by employing the K-means algorithm to group the cells with the most similar HO behavior into a cluster. Next, the future HOs were forecasted, and abnormal HOs were identified. The main target of the proposal is to minimize the signaling load by avoiding the abnormal HOs.

Now we will present some of the research work done on specific machine learning algorithms:

##### *a) Support Vector Machine*

Authors in [62] capitalize on Support Vector Machine (SVM) to predict the user location in the next 5 seconds. A framework to minimize HO delay using mobility prediction is proposed. However, they did not validate the framework, neither did they compare their work with the existing proposals. In [63], SVM predicts the next cell in a real-time manner, by combining GPS data, short-term Channel State Information (CSI), and long-term HO history. The presented model was applied on a synthetic Manhattan grid scenario. Results show that CSI results in almost 100% better prediction accuracy compared to using HO history alone. Using different shadowing values to represent different terrain and environment can further strengthen the idea practicality.

##### *b) Neural Networks*

Few works in [64] [65] have leveraged neural networks for mobility prediction. The basic idea is to utilize the neural network to learn mobility-based model for every user and then make prediction about the future serving cell. Authors in [64] performed clustering of the input RSS samples through k-means. The clusters and input RSS samples were then fed to a classifying model, where neural network was used to predict the user position. Results show that the prediction accuracy increase by just 5% when compared to the prediction using neural networks alone.

### 3) MARKOV CHAIN BASED PREDICTION

A large number of research studies have used Markov chain based approaches for mobility prediction for their ability to yield better accuracy than most other predictors with lower complexity [66]. In the following, we review recent studies for commonly used Markov Chain (MC) variants:

#### a) *Standard Markov Chain:*

Standard Markov Chain is a memory-less algorithm as the next state depends only on the current state and not on the sequence of the events that preceded it.

Authors in [67] extracted trajectories of 4,914 individuals using 27-day log of the mobile network traffic data. They compared the original Markov algorithm with the Lempel-Ziv (LZ) family algorithm [68]. The core operation of the LZ predictor is by maintaining a prediction tree which adds more complexity compared to Markov. It was concluded that although slightly more accurate, LZ family algorithm consumes a lot more resources and time than Markov algorithm. Most of the mobility prediction algorithms only consider spatial factors to predict future movements. Authors in [68] improved Markov Chain based model by adding a temporal factor and achieved 6% higher accuracy.

Humans usually follow regular paths as discussed earlier, however, they may deviate from their accustomed routine at some instances. Authors in [69] proposed a practical model based on State Based Prediction (SBP) method to predict the place to be visited when the user's trajectory exhibits unexpected irregularities. When user diverts from the routine, SBP is employed to conduct the prediction. Experiments reveal that the accuracy of proposed model can reach more than 83%, which is higher than the accuracy of 60% achieved by LZ predictor used in [68].

Authors in [70] proposed an implementation architecture for the MOBaaS (Mobility and Bandwidth prediction as a Service). The MOBaaS can be readily integrated with any other virtualized LTE component to provide the prediction information. Spatial information (location history) and temporal information (time and day data) are collected and analyzed. The results show a 33% reduction in access time for the requested content using the MOBaaS prediction information can be achieved. Due to its appeal, several extensions of MOBaaS were proposed later. For example, in [71], authors stressed that MOBaaS can be implemented in a cloud based mobile network architecture and can be used as a support service by any other virtualized mobile network service. Authors also evaluated the feasibility and effectiveness of the proposed architecture.

Fazio et al. [72] propose Distributed Prediction with Bandwidth Management Algorithm (DPBMA). The algorithm uses Markov Chains to predict the user movement at each BS in a distributed way. This makes the proposed solution different from many other studies [67], [69], [70] where Markov chains are used to improve system utilization by reserving resources prior to the HO. This helps in preventing the call drop occurrences. However, distributed algorithm means BS needs to do a lot of processing making this solution not an attractive option for low cost BS or small-cells.

#### b) *Enhanced-Markov Chain*

In [73], subscriber's mobility is predicted using the enhanced Markov chain algorithm. The core idea is to add the behavior pattern and temporal data of the users from CDR into the Local Prediction Algorithm (LPA) and the Global Prediction Algorithm (GPA). LPA and GPA are based on first and second order Markov processes where transition probability to next cell depends only on the present cell, and both present and previous cell respectively. Results show that the proposed prediction methodology achieves prediction accuracy of 96% compared to GPA with prediction accuracy of 81.5%. However, users without any historical record in the training process showed poor prediction accuracy. Techniques such as particle filter or Kalman filter can be employed to increase accuracy for new users.

#### c) *Semi-Markov Model*

Authors in [74] argue that both discrete and spatial Markov Chain assume human mobility as memory less. By using these approaches, we can achieve spatial prediction of future cell, but time factor cannot be incorporated. To address this concern, authors predicted HO to the neighboring BS using Semi-Markov Model. Semi Markov process allows for arbitrarily distributed sojourn times. Experimental evaluation leveraging on the real network traces generated by the smartphone application showed prediction accuracy of 50% to 90%. An extension of this approach can be to have ping-pong HO predictions.

#### d) *Hidden-Markov Model (HMM)*

Ahlam et al. [75] proposed HO decision algorithm (OHMP) using HMM predictor to accurately estimate the next femto-cell using a) the current and historical movement information, and b) the strength of the received signals of the nearby BSs. The performance of OHMP is validated by comparison with the nearest-neighbor and random BS selection strategies. Results show that the number of ping-pong HOs reduce by 7 times when considering dense deployment of femto cells. Results in [75] are demonstrated for a single user scenario only and does not portray futuristic cellular networks with large number of users. To address this concern, same set of authors extended their idea in [76] by incorporating multiple UEs. They take into consideration the available BS resources of serving femto-cell and interference level from the target femto-cell. The presented OHMP-CAC algorithm introduced a proactive HO scenario where HO is triggered when SINR of the serving cell reaches a predefined threshold. OHMP-CAC minimized the number of HOs by 64% and reduced the average HO decision delay by up to 75% when compared with the traditional RSSI based scheme.

As discussed earlier, mobility prediction using Markov chain is a memory-less system as future state can only be determined by the current state. On the other hand, enhanced Markov Chains are based on historical data, but their application is very complex. Moreover, mobile operators may not be allowed by the customers to use their historical data due to privacy concerns. Even if historical records are accessible, HO delay might still be observed due to the extraction and processing complexity of historical records. Due to these factors, history-based prediction algorithms might render impractical.

### B. Measurement Based Prediction

Measurement based mobility prediction approaches are more accurate than history-based mobility prediction schemes. However, the processing complexity due to the measurement procedure cannot be ignored.

#### 1) RSSI BASED PREDICTION

Soh and Kim [77] introduced RSSI based mobility prediction while keeping in view different UE velocities. They incorporated UE trajectory and road topology information to yield better prediction accuracy. The prediction goal is to achieve timely HO and limit the probability of forced termination during HOs. In addition, bandwidth reservation scheme was proposed that dynamically reserves radio resources at both participating BSs during the HO procedure. Results show that proposed mobility prediction scheme helps achieve almost similar forced termination probability as the benchmark scheme with perfect knowledge of the mobile UE's next cell and HO time.

Authors in [78] proposed an RSSI-based prediction scheme to reduce VoLTE end-to-end delay and HO delay under different UE velocities in mixed femto-cell and macro-cell environments. The core idea is to send the measurement reports based on user velocity and predict when and where to trigger HO procedure. As a result, HO delay is reduced by 28%. For ultra-dense BS deployment, mobile UE may not perform HO to each BS on its trajectory. Future work can include the consideration of load condition, so that both low latency and adequate resources can be guaranteed for improved QoE.

The decision to skip the HO to a better radio condition cell can be based on dwell time or cell load condition. Next femto-cell prediction based on radio connection quality and cell load status is presented in [79]. Authors proposed two cell selection methods; a) BS prediction after analyzing the collected data of average RSSI from nearby femtocells, b) using cognitive radio to sense neighboring femtocells load before triggering HO. Results show that appreciable number of HOs can be avoided when compared with only RSS based HO approach. Thus, data interruption during HO and chances of Radio Link Failure e.g., due to ping-pong HOs can be avoided.

Authors in [80] argue that RSS alone should not be considered when performing inter-RAT HO. Instead current RSS predicted RSS and available bandwidth should be considered. They proposed Fuzzy logic based Normalized Quantitative Decision (FNQD) scheme which aids in eliminating ping-pong effects in HetNets. This work can help realize improved mobility management for LTE-Unlicensed (LTE-U). However, the key performance metrics such as throughput and HO delay should be added for validation purposes.

#### 2) MEASUREMENT REPORT BASED PREDICTION

Song et al. used Grey system theory in [81] to predict the $(N+1)^{th}$ measurement report (MR) from $N^{th}$ MR for high speed railways. The key idea is to utilize the predicted MR to make proactive HO trigger decisions. Their findings showed that the difference between predicted MR and actual MR is within 1%. Thus, the proposed scheme is capable of proactively triggering HO in advance and HO success probability is enhanced from 5% to 10%.

#### 3) USER DIRECTION BASED PREDICTION

Authors in [82] present a user mobility prediction method for ultra-dense networks using Lagrange's interpolation. They predicted user's arrival into their neighboring femtocells based on users moving direction and the distance between users and neighboring cells. The presented approach increases the prediction accuracy when compared with only distance based and direction based mobility prediction. However, the performance of their proposed prediction scheme is not compared with other existing schemes to quantify the performance gains.

#### 4) USER VELOCITY BASED PREDICTION

Higher UE velocity imposes additional threat to reliability making prediction of UE velocity extremely important to help tune the parameters more effectively. 3GPP based solution assigns mobility states (high, medium, low) depending on certain number of HOs in a moving time window. However, this technique will be inefficient in 5G networks with unplanned and highly dense deployment of heterogeneous BS having variable cell radius. UE velocity was estimated in [83] based on the sojourn time sample and accuracy was analyzed via Cramer Rao Lower bound. Numerical results show that the velocity prediction error decreases with the increase in BS density. The authors in [83] further extended their idea in [84]. The predicted UE velocity was used to assign the appropriate mobility state. Validation was done by gathering statistics of the number of HOs as a function of UE velocity, small-cell density, and HO count measurement time window. The results show similar conclusion as in [83] that the accuracy of a suitable mobility state detection (known from UE velocity) increases with increasing small-cell density.

Authors in [85] observed that mobility in urban areas depends on the traffic laws and is affected by the behavior of other people (red signal, other driver brakes etc.). They predicted user mobility based on the observation that a UE with constant velocity will probably go straight, while a UE decreasing in velocity might indicate stoppage on red light or a turn to a different direction. User location in their model is estimated from uplink time difference of arrival or provided by the UE via AGPS while velocity estimation is achieved by increasing sampling rate of location or by Doppler shift. Results showed that overall throughput can be enhanced by 39%, 31%, and 19% for UE velocities ranging from 25, 50, 75 km/h respectively.

TABLE VII
MOBILITY PREDICTION APPROACHES AND THEIR KEY-GOALS

| References | History Based Prediction | | | Measurement Based Prediction | | | | Location Based Prediction | Primary Target KPI | | | | | |
|---|---|---|---|---|---|---|---|---|---|---|---|---|---|---|
| | Cell Trace | Machine Learning | Markov Chain | RSSI | MR | User Direction | User Velocity | | Ping-Pong/ RLF Avoidance | Call Drop Avoidance | HO Delay Reduction | Data Interruption Reduction | Resources Allocation Prior to HO | Prediction Accuracy |
| [60] | ✓ | | | | | | | | | | | | | ✓ |
| [62] | | ✓ | | | | | | | | | | | | ✓ |
| [63] | | ✓ | | | | | | | | | | | | ✓ |
| [64] | | ✓ | | ✓ | | | | | | | | | | ✓ |
| [65] | | ✓ | | | | | | | ✓ | | | | | |
| [67] | | | ✓ | | | | | | | | | | | ✓ |
| [111] | | | ✓ | | | | | | | | | ✓ | | |
| [68] | | | ✓ | | | | | | | | | | | ✓ |
| [69] | | | ✓ | | | | | | | | | | | ✓ |
| [70] | | | ✓ | | | | | | | | | ✓ | | |
| [71] | | | ✓ | | | | | | | | | ✓ | | |
| [72] | | | ✓ | | | | | | | ✓ | | ✓ | ✓ | |
| [73] | | | ✓ | | | | | | | | | | | ✓ |
| [74] | | | ✓ | | | | | | | | | | | ✓ |
| [75] | | | ✓ | ✓ | | | | | ✓ | | | | | |
| [76] | | | ✓ | | | | | | ✓ | | ✓ | | ✓ | |
| [77] | | | | ✓ | | | ✓ | | | | | | ✓ | |
| [78] | | | | ✓ | | | ✓ | | | | ✓ | | | |
| [79] | | | | ✓ | | | | | ✓ | | | | ✓ | |
| [80] | | | | ✓ | | | | | ✓ | | | | | |
| [81] | | | | | ✓ | | | | | | ✓ | ✓ | | |
| [82] | | | | | | ✓ | | | | | | | | ✓ |
| [83] | | | | | | | ✓ | | ✓ | ✓ | | ✓ | | |
| [84] | | | | | | | ✓ | | ✓ | ✓ | | ✓ | | |
| [85] | | | | | | | ✓ | | ✓ | ✓ | | ✓ | | |
| [86] | | | | | ✓ | | | ✓ | | ✓ | | | ✓ | |
| [87] | | | | | ✓ | | | ✓ | | ✓ | | | ✓ | |

### C. Location Based Prediction

The knowledge of UE location can assist in an improved mobility prediction. Effective localization when combined with the mobility prediction algorithms can yield more efficient HO related QoE results.

Soh and Kim in [86] presented a decentralized Road Topology Based mobility prediction technique where the GPS equipped UEs shall perform mobility prediction based on approximated cell boundary data that was shared by the serving BS. Cell boundary data is represented by a set of points at the cell edge and is populated based on historical measurement reports sent by UEs. UE at the cell edge will thus report the corresponding location ID back to the BS, and proactive resource reservation at potential BS can be achieved. Results show considerable reduction in forced termination compared to a reactive HO approach without mobility prediction. This approach can be applied to the macro-cells but is not reasonable to small-cells as mobile UEs will have to send a lot of high-powered uplink messages at cell edge (high path loss condition). This can lead to an increase in HO failure due to high uplink RSSI. Moreover, UE battery consumption will be high.

Authors in [86] proposed mobility prediction scheme based on road topology information. The main idea is based on the approximated cell boundary based on prior HO instances, being configured by the serving cell. The authors in [86] extended their idea in [87] to add the temporal component to mobility prediction. The scheme uses linear extrapolation from a UE positioning data to predict its HO cell and time. 70% mobility prediction accuracy was achieved compared to 60% in their prior work [86].

Location based mobility prediction approaches assume all cell phones to have an accurate position information, which cannot always be guaranteed. Moreover, security concerns of the subscribers may hinder the collection of necessary data to realize accurate cell boundaries.

While proactive mobility management seems to be a great fit to address the stringent QoE requirements in the emerging cellular networks, the trivial network dimensioning tasks should be planned while keeping in view the effect of mobility on the deployed network.

## V. MOBILITY ORIENTED NETWORK PLANNING AND OPTIMIZATION

Realizing massive potential of network densification to address the capacity crunch has introduced additional network planning challenges as discussed by Azar et al. in [88]. One such challenge will be faced due to larger fraction of the mobile users in the network; hence, the network must be planned while considering mobility management in mind. Suitable network architecture can help achieve QoS goals while keeping the cost (e.g. signaling) to a minimum, and ultimately help attain higher network efficiency.

### A. Signaling Minimization by Reduction in Handovers in High Speed Trains

Since considerable signaling overhead is being generated due to a single HO, network planning and architecture aimed to reduce the number of HOs can certainly be very effective. High speed train users are subjected to frequent HO as they move along the track. Apart from a huge amount of signaling data generation, they can also encounter severe issues like RACH failure, late HO, Radio Link Failure (RLF), and Release with Redirect (RwR). Futuristic mobile networks with smaller footprint small-cells will cast an even bigger risk.

To address this problem, authors in [89] presented a HO minimization technique where they propose to install an antenna on top of the train that will perform connectivity and trigger HO with covering BSs. Network deployment approach has been demonstrated in Fig. 11. This elevated antenna interfaces with an inner-train network to serve the passengers. Thus, instead of several users performing HOs simultaneously, only one HO will be performed by the elevated antenna. This not only reduces signaling load, but also minimizes the risk of HO failure as UEs will not experience penetration loss of 20-30 dB inside the train. Field trial conducted on a 2.4km run showed downlink throughput of 1.25Gbps.

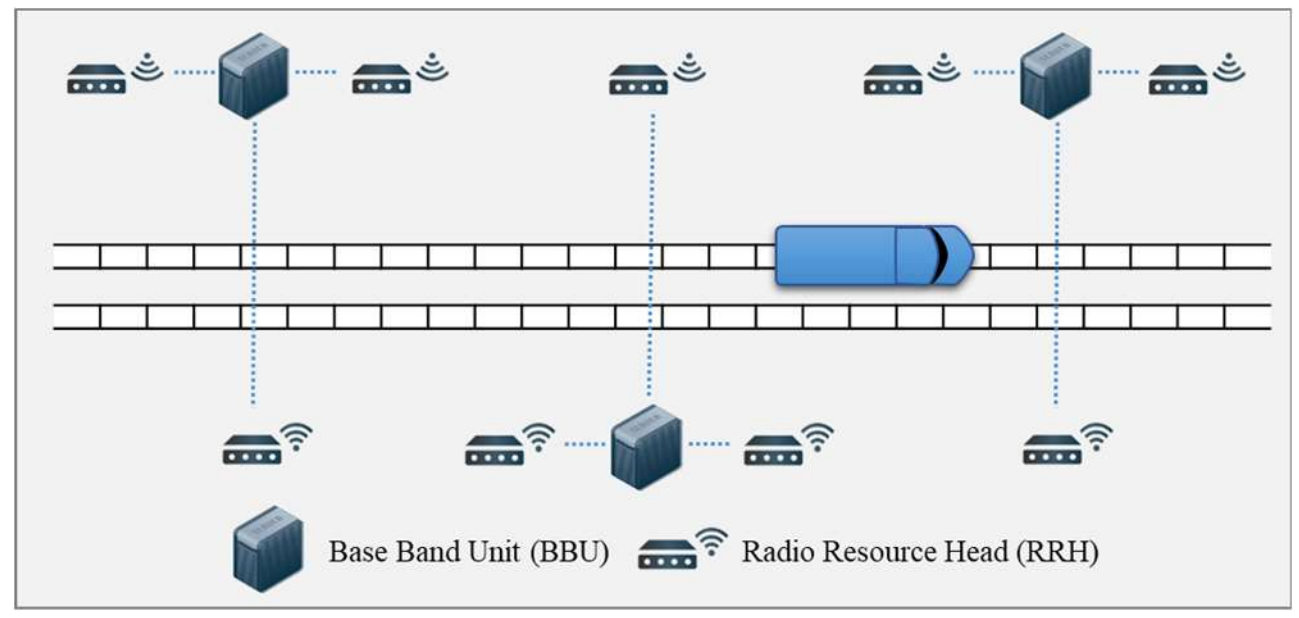

FIGURE 11. **Directional network deployment using RRHs** [89].

The concept of elevated antenna seems practical and is studied even by 3GPP [90]. However, single point of failure lies on its very foundation; if elevated antenna fails and observes HO failure then the multiple users being served under that antenna will have disrupted data transmission. Intelligent switching of the elevated antennas based on proximity to the BS can not only avoid HO failure but also deliver high throughput due to better SINR, but at the cost of complexity and cost. Another drawback will be the latency due to the addition hop between the top-mounted antenna and the inside-train UEs. As a result, self-driven trains in the near future might not achieve the required latency QoE goal.

### B. Changing Core Network (CN) to Achieve Latency Goals

Authors in [11] studied the latency, HO execution time, and coverage of four live LTE networks based on 19,000 km of drive tests. The test was conducted in a mixture of rural, suburban, and urban environments. Their measurements reveal that the lion's share of latency comes from the core network rather than the air interface. Based on the study in [11], Johanna et al. [91] proposed a new entity called the edge node that integrates MME and control plane part of SGW and PGW. Each edge node covers several BS, and when UE moves to coverage of another edge node, the application server and gateway is also shifted to minimize the latency. This approach helps to reduce latency for every HO done within BSs connected to the same edge node. However, HO associated with inter-edge node is followed by IP address reassignment and application-server transfer, which adds to delay and data interruption.

Keeping in view that the number of 5G subscriptions will be 2.6 billion by the year 2025 [1], authors in [5] suggested a simplified 5G core network which will be connectionless, and will incorporate the best effort without the support for node mobility. The core idea is to have a legacy internet-like core network that will not be QoS centric, and the majority of the traffic will flow through default bearers only. Experiments were conducted on a smartphone to show that video streaming, web browsing, and messaging will work well, thus, the future core network can be radically simplified, resulting in a cost-effective solution. The authors in [5] mainly focused on a simplified core network with low complexity. Over-simplification of core network is not a practical approach as major functionalities of billing and access control cannot proceed. Similarly, IP re-allocation at every single HO is not feasible and may result in high latency or even packet loss.

### *C. C/U Plane Split*

With improvement and advancement in the hardware technology, telecom operators can benefit from decoupling control and user plane (see Fig. 12). By doing so, future mobile networks with the composite of macro-cells and small-cells can be used intelligently for efficient resource utilization. Moreover, signaling overhead from large number of HOs can be minimized by assigning control plane and user plane to macro-cells and small-cells respectively.

Authors in [92] address mobility support for high density, flexible deployment of small-cell architecture with flexible backhaul using Localized Mobility Management (LMM) technique. The first step centralizes control-plane from small-cells to a Local Access Server (LAS). The second step allows individual small-cells to handle the mobility events, but still requires the LAS to act as a mobility anchor. Analytical model based on discrete time Markov chain is used to evaluate the average HO signaling cost, average packet delivery cost, average HO latency and average signaling load to the core network. Results show that average HO latency decrease by 10ms compared to the 3GPP scheme [11].

Authors in [93] minimized signaling overhead in a 5G network with a high density of mmWave BSs serving users under the umbrella of macro BSs. C/U split was employed where macro BS provides the control plane and several mmWave cells were group into clusters. Inter-cell HO signaling was curtailed by using a gateway cluster controller, resulting in signaling reduction in the core network as well. Results show that normalized X2 signaling overhead reduces from 100% to 10% as the density of the deployed mmWave cells increases.

Authors in [94] targeted latency minimization in their proposed novel mobility management scheme for intro-domain handover (HO within the same SDN domain) and inter-domain handover (HO across different SDN domains). Layer 2 information and buffering approach was used to achieve HO latency of just 400ms compared to the legacy DMM with 100ms of HO latency.

While proactive mobility management and mobility-oriented network planning seem to deliver promising results, the constant temporal variations in a live network and the importance of key landmarks can be addressed by introducing Artificial Intelligence (AI) to the cellular network domain.

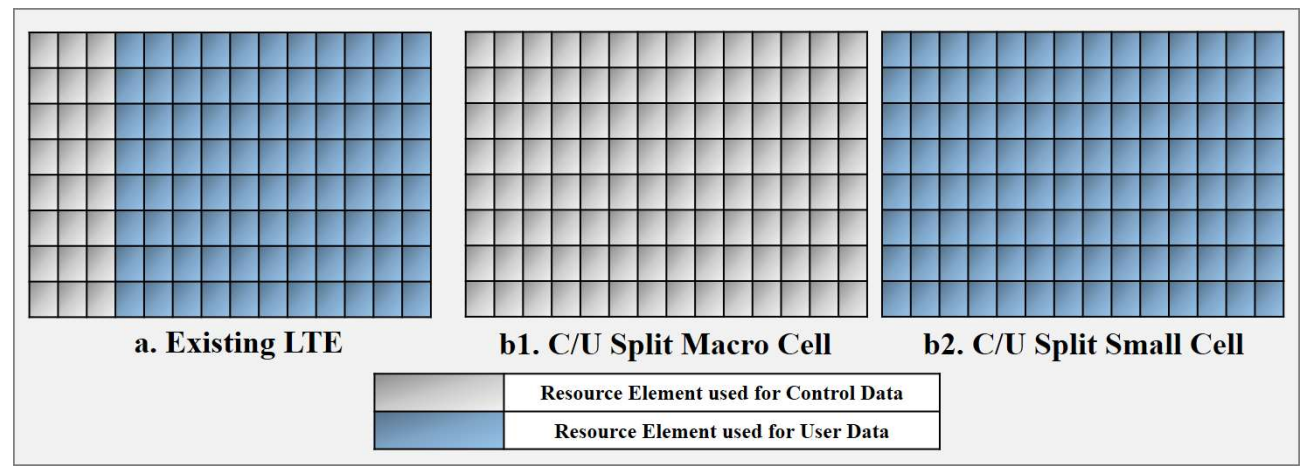

FIGURE 12. **Frame structure for legacy LTE vs C/U plane split architecture.**

## VI. AI-ASSISTED MOBILITY MANAGEMENT

In recent years, AI has gained much popularity for proactively managing mobility in future cellular networks. This is primarily because of an increasing number of configuration parameters and due to the complex interaction between network parameters and associated KPIs (as illustrated in Fig. 8). Once the research community is able to overcome those complex challenges, AI-assisted solutions will have a revolutionary effect on the telecom industry. The tutorial section (Section II) of this paper can help researchers understand the convoluted interplay between the network parameters and affected KPIs. Now we will present some of the AI enabled mobility management solutions present in the literature. The comparison of the presented algorithms can be found in Table VIII.

The mobility prediction algorithm is presented in [95]. Authors use realistic mobility patterns to capture the human movement and a 3GPP compliant 5G simulator was used to represent the HetNets scenario. Results show that mobility prediction accuracy of almost 87% can be achieved for 2dB shadowing with XGBoost compared to 78% with Deep Neural Network (DNN). The work can be extended by using time series predictors such as recurrent neural network or LSTM.

Authors in [96] employed XGBoost supervised machine learning algorithm to perform partially blind HOs from sub-6GHz to co-located mmWave cell. Authors show that this machine learning-based algorithm to achieve partially blind HOs can improve the HO success rate in a realistic network setup of co-located cells. The proposed algorithm should be compared with the existing HO approach in terms of energy efficiency and RLF to further validate the efficacy of the algorithm.

The idea of inter-frequency HO from a macro-cell to a non-co-located high frequency cell with a much lower footprint is presented in [97]. The authors use the Random Forest classification approach and also presented a use case of load balancing by which an efficient resource utilization for the static users can be achieved. The shortcoming in the presented approach is that for high-speed users, the load balancing based HO to smaller footprint cell may be inefficient due to large HO rate and the resultant signaling overhead and chances of HO failure.

Authors in [98] develop a Reinforcement Learning (RL) based HO decision algorithm for the mmWave cells by taking into account the user experience as a weighted sum of throughput and HO cost. Based on the user's mobility information, the optimal beamwidth is selected by considering the trade-off between the a) directivity gain and b) beamforming misalignment. The algorithm approves the HO trigger for mobile users depending on UE velocity and BS density. The work can be extended by evaluating the signaling overhead reduction and throughput gain achieved when compared with other existing algorithms in the literature.

TABLE VIII
AI-ASSISTED MOBILITY MANAGEMENT APPROACHES

| References | AI Tool Used | mmWave Support | Real Network Data | Wrong HO Elimination | User Velocity as Input | Realistic Mobility Patterns | HO Anomaly Detection | HO Parameter Optimization | Load Balancing Support | Key Objective |
|---|---|---|---|---|---|---|---|---|---|---|
| [95] | XGBoost<br>DNN | | | | ✓ | ✓ | | | | User Mobility Prediction |
| [96] | XGBoost | ✓ | | | | | | | | Semi Blind HO from sub-6GHz to co-located mmWave BS |
| [97] | Random Forest | ✓ | | | | | | | ✓ | HO to non-co-located mmWave BS |
| [98] | Reinforcement Learning | ✓ | | ✓ | | | | | | QoE aware HO to mmWave cell |
| [99] | LSTM / RNN | | | ✓ | ✓ | | | | | Future RSRP prediction and RLF avoidance |
| [101] | Neural Networks | | | ✓ | | | | | | HO prevention to BSs with coverage holes |
| [102] | K-Means<br>Gaussian Processes | | ✓ | | | | ✓ | | | HO performance monitoring through KPI data |

Authors in [99] predicted the RSRP of the serving and the HO target cell using Long Short-Term Memory (LSTM) and Recurrent Neural Network (RNN). The algorithm also predicts RLF instances with an accuracy of 84% using only RSRP as an input feature. An extension to [99] has been made in [100] where other features like SINR, out-of-sync identifier, RACH issues, and max RLC retransmission have been used for RLF prediction.

A wrong HO avoidance algorithm has been proposed in [101]. It uses neural networks to prevent the HO to BSs which are affected by the undesirable radio propagation scenarios in the network, e.g., coverage hole caused by an obstacle. The proposed algorithm enables a UE to learn from past experiences (coverage unavailability) to select the best cell for HO in terms of QoE. The authors show that their algorithm helps achieve users to successfully complete the downlink transmissions more than 93% of the time. However, the simulation environment is quite simplistic where the UE traverses a straight line with only three BSs along the way. Hence, the movement of UE is almost deterministic, and the Neural Network can easily learn its pattern and can identify the optimal BS to perform HO. Furthermore, a single test UE gives a limited evaluation of the proposed algorithm. Elaborated results with a HetNet scenario and arbitrary movement of multiple users will have more realistic results.

Based on HO attempts per hour, authors in [102] cluster cells into different groups with similar HO profiles using the K-means algorithm. For each cluster, hourly HO attempts were forecasted using linear regression, polynomial regression, neural networks and gaussian processes. the highest R2 value of 0.99 was obtained when using the gaussian process. The proposed model then checks for abnormal HO behavior e.g. ping-pong. Future work can be to proactively predict abnormal HO behavior ahead of time and to recommend suitable proposed parameters to prevent HO KPI degradation.

## VII. FUTURE RESEARCH DIRECTIONS AND CONCLUDING REMARKS

Ultra-Dense Cellular Networks (UDN) containing mmWave based small-cells are being considered an essential part of the future vision of cellular systems vis-à-vis 5G and beyond. Harnessing mmWave spectrum has a strong potential to solve the two long-standing problems in cellular networks: spectrum scarcity and interference. Remarkably, most research towards UDN remains focused on channel modelling and hardware design aspects of the mmWave based UDN, and mobility management in UDN so far remains a Terra incognita. The panorama of mobility challenges arising in emerging mobile networks implies that if no drastic and timely measures are taken to rethink mobility management for future UDN, user mobility management can become the bottleneck in practical deployments of UDN despite advances in the hardware design of mmWave and conventional spectrum based small-cells. Enabling seamless mobility in futuristic mobile networks require much complex network design and planning in order to achieve the QoE goals and to address the intricacies of the network architecture needed to realize the promised user experience. The high throughput requirement, heterogeneity of UEs and BSs, and security awareness of 5G environments appeal for a fast, distributed and privacy preserved mobility management. This article provides an extensive survey of mobility management for future cellular networks. As studied in the prior section, researchers have added healthy contributions in an attempt to realize an optimal and satisfactory network. However still, some research domains are untouched or haven't been given the attention they deserve. Now we will discuss a few of the key points related to future research directions:

### A. HO Delay Based SINR Distribution

Current SINR modelling is based on best-server-association, however, the UE always camp on the second-best cell prior to HO. This is the result of the HO evaluation process [18] which ensures that the target cell is the best candidate cell for HO. A mobility oriented SINR distribution which captures the temporal negative SINR [103] before HO needs to be studied for more realistic throughput estimation.

### B. HO Delay Based Uplink Interference

Current researchers do not consider the practical situation where due to intra-frequency HO delay, high mobility users are closer to the target cell while still being served by the comparatively farther located serving cell. Under those circumstances, high uplink power to achieve target SINR in the serving cell can cause strong temporal interference in the target cell. The issue can be aggravated under highly dense BSs deployed in an impromptu fashion. However, this problem can be tackled by utilizing an eICIC ABS (Almost Blank Subframe) scheme for highly mobile users. Proactive HO trigger can also eliminate the possibility of high uplink RSSI by performing timely HO.

### C. Latency Goals

Another challenging aspect of the small cell deployment is that the small-cells are typically not directly connected to the core network and lack Xn or N2 interfaces (for inter-cell communication) which are the real means of coordinating mobility procedures in the macro-cells. The lack of a low latency connection to the core network can contribute to significant HO signaling delays.

### D. Energy-Efficiency

Achieving both UE and network-level energy efficiency is a big challenge for futuristic cellular networks, especially when considering ultra-dense BS deployment and the addition of a wide variety of user devices. Most of the existing energy-saving schemes have a common tenancy; cells are switched ON/OFF reactively in response to changing cell loads. A meritorious effort has been made by Hasan et al. in [57], where authors proposed the AURORA framework in which the past HO traces are utilized to determine future cell loads. The prediction is then used to proactively schedule small-cell sleep cycles. Load balancing is also achieved through the use of appropriate Cell Individual Offset (CIO).

### E. Smart Intra-Frequency Search

Dense deployment poses challenges for small-cell discovery as conventional cellular networks broadcast a neighbor list for the user to learn where to search for potential HO cells. However, such a HO protocol does not scale to the large numbers of neighboring small-cells and the underlying network equipment is not designed to rapidly change the neighbor cell lists as small-cells come and go.

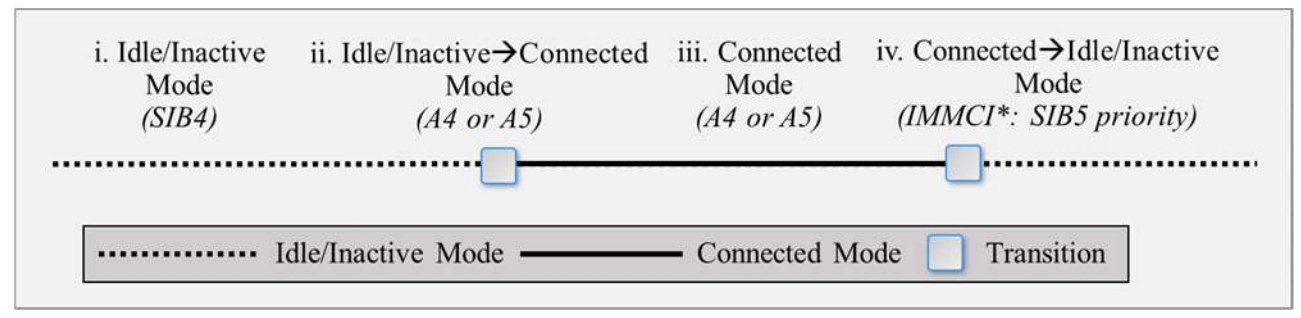

FIGURE 13. **Load Balance (LB) opportunities (i, ii, iii, iv) in different stages of 5G UE connection.**

### F. Smart Inter-Frequency Search

Inter-Frequency (IF) mobility is a vital component of cellular networks but has not got the attention it deserved in the research community. IF-mobility requires event A2 to be triggered, which is followed by the BS to configure measurement gap periodicity to the UE. However, this process interrupts data transmission and reception. This is because UE shifts the radio to measure appropriate IF-cell(s). Futuristic mobile networks with a variety of frequencies ranging from HF to mmWave band may require the UE to undergo an extensive search of available frequencies before initiating a mobility decision. This issue can be aggravated when considering the latency goal of <1ms.

### G. Improving Mobility Load Balancing

Mobility Load Balance (MLB) is a vital component of heterogeneous multi-layer cellular networks and are open to the following challenges:

- LB can be achieved at four different instances as shown in Fig. 13. It can be triggered through i) idle mode SIB4 configuration, ii) after network access using A4 or A5 measurement report, iii) in connected mode using A4 or A5 measurement report (as configured), iv) when UE is released from connected to idle mode using 3GPP proposed IMMCI (Idle Mode Mobility Control Info). In IMMCI, traffic steering is achieved by varying the idle mode SIB5 priority of the serving or target layer. LB in idle mode is the most optimal as signaling and data interruption associated with connected mode LB can be avoided. Moreover, complexity in parameter configuration and management by IMMCI can be minimized. Research contributions are currently lacking for idle mode load balancing. Similarly, a new variant of IMMCI (SON based) is needed which can adaptively steer traffic to achieve load balancing under varying load conditions.
- LB detail procedure has not been provided by 3GPP and is left intentionally to vendors for innovation purposes. LB requires the exchange of load information between participating BSs via the Xn interface. However, different vendors have their own proprietary version of LB implementation, thus, inter-vendor BS cannot perform LB due to mismatch in LB metrics. The existing LTE networks deploy offloading feature, where high load cell offload users to another vendor cell without considering its load condition. This can cause service rejection and ping-pong

HO conditions. The frequent IF-search will disrupt continuous reception and will result in higher latency. 5G heterogeneous network can assume numerous vendors, and to benefit from the load balancing feature, a standard inter-vendor LB mechanism need to be devised.
- Cells with smaller footprints will have few serving UEs, and mobility-based ingress and egress of even a single user can have drastic load imbalance among available frequency bands. Hence, ways to achieve proactive LB is mandatory to have fairness and efficient resource utilization.

### H. Mobility in mmWave Networks

mmWave with bandwidth as large as 500MHz is the remedy to the spectrum saturation in the HF band, however, an intrinsic feature of narrow beams can pose serious challenges in supporting mobility in the emerging cellular networks. Few of the main challenges are presented here:

- Simic et al. [104] practically demonstrates mmWave to prove multi-Gbps connectivity but conclude that supporting mobility is a very challenging task due to the outage area of as high as 40% with 90BS/km$^2$ deployment. The reason for the coverage hole is the high diffraction phenomena in mmWaves, and absence of Non- Line of Sight (NLoS) paths.
- Corner Effect: Indoor areas have cell edge near doors, where the user is more likely to make a sharp turn and hence, time available for HO would be very less especially in the 60GHz mmWave scenario. This issue suggests that some sophisticated techniques, other than conventional methods are required for the HO trigger.
- Current mmWave standards such as IEEE 802.11ad follows the max-RSSI based approach for UE-BS association, however, this solution appears rudimentary and ineffective for emerging network with an ultra-dense BS density. There will be chances of an unbalanced number of users per BS, and ping-pong HOs will be highly likely.
- In addition, cell discovery for mobile users is a major challenge due to the absence of Reference Signal (RS) broadcast as in HF bands.

Presently, an overwhelming understanding of the research circle is to use mmWave-cells for static users only. Intricacies of mobility between the beams (of both intra-frequency cells and inter-frequency cells) need to be addressed to support mobility. One possible solution is to come up with a hybrid solution where HF macro-cells with much accurate UE location guide the UEs how, when and to which small-cell they need to connect to. This is similar to control-data split architecture with mmWave providing data support while UE is under the coverage of macro-cell providing control signals.

### I. Low-Cost Multi-Connectivity

Dual connectivity architecture has been proposed to mitigate mobility management problems in HetNets by allowing UE to connect with the macro-cell for control connectivity as well as simultaneous data connectivity with small-cells.

The effect of the user association on dual connectivity performance is an interesting research problem that needs to be investigated in detail. Researchers need to study the gain dual connectivity can yield in terms of HO overhead reduction, synchronization complexity, and radio resource efficiency.

Most of the research work address reliability and latency goals through multi-connectivity, however, signaling load increment is not addressed. More efficient proposals with special consideration of signaling load need to be devised.

### J. Accurate and Efficient Mobility Prediction

The mobility prediction schemes are seen as a driving force for context aware cellular network as they are used to proactively reserve resources, trigger LB, and activate/deactivate small-cells. Few challenges associated with mobility prediction are:

- Users not willing to share location information due to the privacy reasons.
- GPS data acquisition consume user battery and intermittent accessibility requests resulting in signaling or RACH issues (some RACH failure issues cannot be seen in the KPI data).
- Accuracy and reliability of 3GPP proposed Minimization of Drive Test (MDT) feature is needed to be evaluated since multitude of factors like the GPS error [105], quantization resolution etc. affect the accuracy of the measurements reported by the UE.
- Although human trajectory exhibits high predictable component [56], however, mobility prediction is always bound to have some inaccuracy as can be understood through an example: an office employee may have lunch in a canteen, in a conference room, with colleagues in an outside restaurant etc. These random variations are almost impossible to predict.

A possible solution can be resource reservation to be done in the multiple neighbors, however, the cost of signaling and available resource for other UEs especially during busy hour needs to be considered.

## ACKNOWLEDGMENT

This work is supported by the National Science Foundation under Grant Numbers 1718956 and 1559483 and Qatar National Research Fund (QNRF) under Grant No. NPRP12-S 0311-190302. The statements made herein are solely the responsibility of the authors. For more details about these projects please visit: http://www.ai4networks.com.